\documentclass[preprints,article,accept,moreauthors,pdftex]{Definitions/mdpi}

\firstpage{1} 
\pubvolume{1}
\issuenum{1}
\articlenumber{0}
\pubyear{2022}
\copyrightyear{2020}
\datereceived{} 
\dateaccepted{} 
\datepublished{} 
\hreflink{https://doi.org/} 
\pdfoutput=1

\Title{Detectability of the cross-correlation between CMB lensing and stochastic GW background from compact object mergers}

\TitleCitation{Title}

\Author{Giulia Capurri $^{1,2,3}$*, Andrea Lapi $^{1,2,3,4}$ and Carlo Baccigalupi $^{1,2,3}$}

\AuthorNames{Firstname Lastname, Firstname Lastname and Firstname Lastname}

\AuthorCitation{Capurri, G.; Lapi, A.; Baccigalupi, C.}

\address{%
$^{1}$ \quad SISSA, via Bonomea 265, 34136, Trieste, Italy\\
$^{2}$ \quad INFN-Sezione di Trieste, via Valerio 2, 34127 Trieste, Italy\\
$^{3}$ \quad IFPU, via Beirut 2, 34151, Trieste, Italy\\
$^{4}$ \quad IRa-INAF, Via Gobetti 101, 40129 Bologna, Italy\\}

\corres{Correspondence: giulia.capurri@sissa.it}

\abstract{The anisotropies of the Stochastic Gravitational-Wave Background (SGWB) produced by merging compact binaries constitute a possible new probe of the Large-Scale Structure (LSS). However, the significant shot noise contribution caused by the discreteness of the GW sources and the poor angular resolution of the instruments hamper the detection of the intrinsic anisotropies induced by the LSS. In this work, we investigate the potential of cross-correlating forthcoming high precision measurements of the SGWB energy density and the Cosmic Microwave Background (CMB) lensing convergence to mitigate the effect of shot noise. Combining a detailed model of stellar and galactic astrophysics with a novel framework to distribute the GW emitters in the sky, we compute the auto- and cross-correlation power spectra for the two cosmic fields, evaluate the shot noise contribution and predict the signal-to-noise ratio. The results of our analysis show that the SGWB energy density correlates significantly with the CMB lensing convergence and that the cross-correlation between these two cosmic fields reduces the impact of instrumental and shot noise. Unfortunately, the S/N is not high enough to detect the intrinsic SGWB anisotropies. Nevertheless, a network composed of both present and future generation GW interferometers, operating for at least 10 yrs, should be able to measure the shot noise contribution.}


\newcommand{\omegagw}{\Omega_{\rm{gw}}}
\newcommand{\baromega}{\bar{\Omega}_{\rm{gw}}}
\newcommand{\mc}{\mathcal{M}_{c}}
\newcommand{\fobs}{f_{\rm{o}}}
\newcommand{\eobs}{\boldsymbol{\hat{e}_{\rm{o}}}}

\begin{document}

\section{Introduction}
The direct detection of gravitational waves (GW) \citep{LIGOScientific:2016aoc} achieved by the network of ground-based interferometers constituted by the Advanced Laser Interferometer Gravitational-Wave Observatory (LIGO) \citep{LIGOScientific:2014pky}, Advanced Virgo \citep{VIRGO:2014yos} and the Kamioka Gravitational Wave Detector (KAGRA) \citep{Somiya:2011np} has opened a new observational window on the Universe. The past five years have witnessed a real revolution in astronomy, with the detection of more than 50 compact binary coalescences during the first three LIGO/Virgo/KAGRA observing runs \citep{LIGOScientific:2018mvr,LIGOScientific:2020ibl,LIGOScientific:2021usb}, the last of which is still ongoing. Each of these detections is associated with a single loud event, but we expect that all the GW signals too faint or too numerous to be individually resolved combine together to create a stochastic
gravitational-wave background (SGWB). The measurement and characterization of the SGWB are one of the main observational targets for current and future generation GW detectors. Existing data from the LIGO/Virgo/KAGRA network have already placed upper bounds on both the isotropic and anisotropic components of the SGWB \citep{KAGRA:2021kbb,KAGRA:2021mth}.

Many physical mechanisms, both astrophysical and cosmological, can produce different backgrounds with distinct properties (see \citep{Christensen:2018iqi} for a review). Among them, the SGWB given by the incoherent superposition of GW events produced by merging stellar remnant compact binaries has raised significant interest among the scientific community \citep{Regimbau:2011rp,Rosado:2011kv,Marassi:2011si,Zhu:2011bd,Zhu:2012xw,Wu:2011ac,KowalskaLeszczynska2015,TheLIGOScientific:2016wyq,Abbott:2017xzg, Perigois:2020ymr}. This astrophysical SGWB is one of the dominant contributions in the frequency range explored by ground-based detectors (Hz-kHz) and is likely to be the first one to be detected. Moreover, the merging compact binaries that create this SGWB are the outcomes of stellar evolution and mainly reside in galaxies. Consequently, the SGWB anisotropies reflect the distribution of galaxies in the Universe and could constitute a new tracer of the Large-Scale Structure (LSS). For these reasons, the anisotropic SGWB has been extensively studied during the last few years, with relevant effort given to theoretical modeling \citep{Contaldi:2016koz,Cusin:2017fwz,Cusin:2017mjm,Jenkins:2018uac,Jenkins:2018kxc,Cusin:2018rsq,Cusin:2019jpv,Bertacca:2019fnt,Pitrou:2019rjz,Capurri:2021zli,Bellomo:2021mer,Cusin:2019jhg}, observational searches \citep{LIGOScientific:2019gaw,LIGOScientific:2016nwa,Renzini:2018nee,Mentasti:2020yyd} and data analysis techniques \citep{Thrane:2009,Gair:2014rwa,Romano:2015uma,Ain:2018zvo,Renzini:2018vkx,Conneely:2018wis,Hotinli:2019tpc}. 

The two main obstacles to the detection of the SGWB anisotropies in the Hz-kHz band are the poor angular resolution of GW detectors to a diffuse SGWB mapping and the presence of a large shot noise contribution. The former issue is related to the noise properties of the detector and how they are mapped onto the sky, but it also depends on the network configuration and the scan strategy \citep{Alonso:2020rar}. Instead, the shot noise arises because the SGWB is composed of discrete transient events occurring at a low rate. Therefore, the predicted SGWB energy density is affected by a high level of uncertainty. Several recent studies have addressed the issue of shot noise, showing that its expected amplitude is orders of magnitude higher than the correlation induced by the LSS \citep{Alonso:2020mva,Jenkins:2019nks,Jenkins:2019uzp}. Cross-correlation with other tracers of the LSS has already been proposed as a possible solution to reduce the impact of shot noise. In particular, cross-correlations with galaxy number counts \citep{Cusin:2017fwz,Cusin:2019jpv,Alonso:2020mva,Canas-Herrera:2019npr,Yang:2020usq,Mukherjee:2019oma}, weak lensing \citep{Cusin:2017fwz,Cusin:2019jpv} and Cosmic Microwave Background (CMB) temperature fluctuations \citep{Ricciardone:2021kel,Braglia:2021fxn} have been investigated. 

Polarization CMB measurements from ongoing \citep{BICEP:2021xfz,ACT:2020goa,POLARBEAR:2019kzz,quijote:2021,SPT-3G:2021eoc} and planned  \citep{Dahal:2021uig,LSPE:2020uos,Ade2019,CMB-S4:2020lpa,Ganga2019European,LiteBIRD:2019} CMB observations are targeting the cosmological background of GWs generated in the Early Universe. Moreover, deflections of CMB photons through weak gravitational lensing caused by forming cosmological structures are explicitly targeted by arcminute scale CMB probes, in both total intensity and polarization, and this aspect is actually central for the present work.

In this paper, for the first time to our knowledge, we cross-correlate the astrophysical SGWB produced by merging compact binaries in galaxies with the CMB lensing convergence \footnote{Although the cross-correlation of galaxy and CMB lensing with resolved GW sources was already employed as a probe of general relativity and dark energy \citep{Mukherjee:2019wfw,Mukherjee:2019wcg}, this is the first time to our knowledge that CMB lensing is cross-correlated with the SGWB, exploiting the fact that both cosmic fields are good tracers of the underlying dark matter distribution.}.The purpose of this study is to verify if the cross-correlation with another cosmic field tracing the same underlying dark matter distribution can bring out the intrinsic SGWB anisotropies given the presence not only of the shot noise but also of the more severe instrumental one. For this reason, we opted for the CMB lensing convergence, which is an integrated tracer of the LSS, just as the SGWB, and will be constrained by future CMB experiments with exquisite precision. Indeed, given the huge uncertainties still affecting GW measurements, it is crucial that the other cosmic field is tightly constrained.
Starting from a detailed model of stellar astrophysics \citep{Chruslinska:2017odi,Chruslinska:2018hrb} and galaxy evolution \citep{Boco:2019teq,Boco:2020pgp}, we compute the anisotropies of the SGWB for three types of sources - binary black holes (BH-BH), binary neutron stars (NS-NS) and black hole-neutron star binaries (BH-NS) - and for different present and future ground-based detectors: LIGO/Virgo/KAGRA, Einstein Telescope (ET) \citep{Sathyaprakash:2012jk} and Cosmic Explorer (CE) \citep{Reitze:2019iox}. 
Exploiting the novel framework to distribute the GW emitters in the sky and simulate a full-sky map of the SGWB developed in \citep{Capurri:2021zli}, we estimate the shot noise contribution.  
Finally, we investigate the potential of cross-correlation with forthcoming high precision measurements of CMB lensing by the Simons Observatory (SO) \citep{Ade2019}, a powerful future probe observing arcminute scale CMB anisotropies with high sensitivity that will allow to characterize the lensing signal with unprecedented precision and to exploit it to enhance the sensitivity on cosmological GWs \citep{Namikawa:2021gyh}. 

The paper is structured as follows: in Section \ref{sec:methods} we go through the theoretical background, in Section \ref{sec:results} we present and discuss our results and in Section \ref{sec:conclusion} we draw our conclusions. In Appendices \ref{app:sgwb_characterization} and \ref{app:shot_noise} we review the characterization of the SGWB produced by merging compact binaries and the derivation of the shot noise contribution, both described in \citep{Capurri:2021zli} and adopted also for this study. 

Throughout this work we assume the standard flat $\Lambda$CDM cosmology with parameter values from the Planck 2018 legacy release \cite{Aghanim:2018eyx}, with Hubble rate today corresponding to $H_{0}= 67.4$ km s$^{-1}$ Mpc$^{-1}$, Cold Dark Matter (CDM) and baryon abundances with respect to the critical density corresponding to $\Omega_{\rm{CDM}}h^{2}= 0.120$ and $\Omega_{b}h^{2}=  0.022$, respectively, reionization optical depth $\tau=0.054$, amplitude and spectral index of primordial scalar perturbations corresponding to ln$(10^{10}A_{S})=3.045$ and $n_{S}=0.965$, respectively.

\section{Methods} \label{sec:methods}

In this Section, we review the description of CMB lensing, closely following \citep{Bianchini2015, Bianchini2016} and references therein. Subsequently, we characterize the SGWB as a tracer of the LSS and we derive an expression of its kernel. Finally, we outline the theoretical background for the cross-correlation of two cosmic fields.

\subsection{CMB lensing} 
The LSS deflects CMB photons during their travel from the last-scattering surface to the observer, leaving small imprints on the observed CMB anisotropies. In particular, the effect of gravitational lensing on CMB photons can be described as a remapping of the unlensed temperature anisotropies $\Theta (\eobs)$ by a two-dimensional vector field in the sky, namely the deflection field $\boldsymbol{d}(\eobs)$:

\begin{equation}
\begin{aligned}
    \tilde{\Theta}(\eobs) & = \Theta ( \eobs + \boldsymbol{d}(\eobs) ) \\
    & = \Theta ( \eobs + \nabla \phi (\eobs))  \\
    & = \Theta(\eobs) + \nabla^{i} \phi (\eobs) \nabla_{i} \Theta(\eobs) + \mathcal{O} (\phi ^ {2}),
\end{aligned}
\end{equation}
where $\tilde{\Theta}(\eobs)$ is the lensed temperature fluctuation and $\phi (\eobs)$ is the CMB lensing potential: 

\begin{equation}
    \phi (\eobs) = -2 \int_{0}^{z_{\star}} \dfrac{c dz}{H(z)} \dfrac{\chi_{\star} - \chi(z)}{\chi_{\star} \chi(z)} \Psi(\chi(z) \eobs, z).
\end{equation}
In the previous equation, $\chi (z)$ is the comoving distance to redshift z, $\chi_{\star}$ is the comoving distance to the last-scattering surface at $ z_{\star} = 1090$, $H(z)$ is the Hubble parameter at redshift $z$, $c$ is the speed of light and $\Psi(\chi(z) \eobs, z)$ is the three-dimensional gravitational potential at the point on the photon path given by $ \chi(z) \eobs$.
The deflection field is defined as $\boldsymbol{d}(\eobs) = \nabla \phi (\eobs)$, where $\nabla$ is the the two-dimensional gradient on the sphere. Since the lensing potential is an integrated measure of the gravitational potential, it is convenient to describe the CMB lensing by means of the lensing convergence, which is proportional to the two dimensional Laplacian of the lensing potential and can be written as a weighted integral over redshift of the projected dark matter density contrast $\delta$:

\begin{equation}
    \kappa (\eobs) =  - \frac{1}{2} \nabla^{2} \phi (\eobs) =  \int_{0}^{z_{*}} W^{\kappa}(z) \delta (\chi(z)\eobs, z).
\end{equation}
The weight inside the integral is the lensing kernel $W^{\kappa}$, which describes the lensing efficiency of the matter distribution and is given by: 

\begin{equation} \label{eq:lensing_kernel}
    W^{\kappa}(z) = \dfrac{3 \Omega_{m}}{2c} \dfrac{H_{0}^{2}}{H(z)} (1+z) \chi(z) \dfrac{\chi_{*} - \chi(z)}{\chi_{\star}}, 
\end{equation} 
where $\Omega_{m}$ and $H_{0}$ are the present-day value of the matter density and the Hubble parameter, respectively.

\subsection{Stochastic Gravitational Wave Background}
The stochastic gravitational wave background is usually described in terms of the dimensionless energy density parameter:

\begin{equation}
\omegagw(\fobs, \eobs ) 
= \dfrac{1}{\rho_{c}} \dfrac{d^{3} \rho_{\rm{gw}}(\fobs, \eobs )}{d \ln \fobs \, d^{2} \omega_{\rm{o}}}
= \dfrac{8 \pi G \fobs}{3H_{0}^{2}c^{2}} \dfrac{d^{3} \rho_{\rm{gw}}(\fobs, \eobs )}{d \fobs \, d^{2} \omega_{\rm{o}}}\,,
\end{equation}
where $\rho_{c} = 3H_{0}^{2}c^{2}/8\pi G$ is the critical density and $\rho_{\rm{gw}}$ is the SGWB energy density at the observed frequency $\fobs$, arriving from a solid angle $\omega_{\rm{o}}$ centred on the observed direction $\eobs$. The energy density parameter can be split into an isotropic term $\baromega (\fobs)$ and a directional dependent term $\delta \omegagw (\fobs, \eobs)$:

\begin{equation}
\omegagw(\fobs, \eobs) =\dfrac{\baromega (\fobs)}{4 \pi} + \delta \omegagw(\fobs, \eobs)\,.
\end{equation}
In this work we consider the astrophysical SGWB given by the incoherent superposition of GW signals produced during the merger of compact binaries inside galaxies. Assuming that the SGWB, as well as the galaxies that host the merging binaries, traces the peaks of the dark matter distribution, the energy density contrast $\delta_{\rm{gw}} =  \delta \omegagw / \baromega $ can be expressed as a line-of-sight integral of the dark matter density contrast: 
\begin{equation}
    \delta_{\rm{gw}} (\fobs, \eobs) = \int_{0}^{z_{\star}} W^{\Omega}(\fobs,z) \delta (\chi(z)\eobs, z), 
\end{equation}
where the SGWB kernel $W^{\Omega}(\fobs,z)$ is the sum of two terms: 

\begin{equation} \label{eq:sgwb_kernel}
    W^{\Omega}(\fobs,z) = \dfrac{b_{\Omega}(\fobs,z)  \dfrac{d\Omega}{dz}(\fobs,z)}{ \biggl( \int dz'  \frac{d\Omega}{dz'} \biggl) } +  \mu(\fobs,z).
\end{equation}
The first term is the product of the linear bias $b_{\Omega}$, which quantifies the mismatch between the distribution of the SGWB and the total matter density, and the SGWB redshift distribution $d\Omega/dz$. The second term takes into account the effect of lensing magnification on the observed SGWB energy density and it is given by: 

\begin{equation}
    \mu(\fobs,z) = \dfrac{3\Omega_{\rm{m}}}{2c} \dfrac{H_0^2}{H(z)}(1+z)  \chi(z) \int_{z}^{z_{\star}} dz' \biggl( 1- \dfrac{\chi(z)}{\chi(z')} \biggl) \bigl(s_{\Omega}(\fobs,z)-1\bigl) \dfrac{\dfrac{d\Omega}{dz}(\fobs,z)}{ \Bigl( \int dz'  \frac{d\Omega}{dz'} \Bigl) }, 
\end{equation}
where $s_{\Omega}$ is the magnification bias. 

The expressions of the redshift distribution $d\Omega/dz$, the bias $b_{\Omega}$ and the magnification bias $s_{\Omega}$ are derived and discussed in \citep{Capurri:2021zli}: we review them in Appendix~\ref{app:sgwb_characterization} for the interested reader. All these quantities can be evaluated once a specific model of stellar astrophysics and galaxy evolution has been chosen: in this work, we adopt the astrophysical prescriptions presented in \citep{Boco:2019teq,Boco:2020pgp}. 
There, the authors combine the results of the \texttt{STARTRACK} population synthesis simulations\footnote{Simulation data publicly available at \url{https://www.syntheticuniverse.org/}}, specifically the `reference B' model in \citep{Chruslinska:2017odi,Chruslinska:2018hrb}, with  different observationally derived prescriptions for the host galaxies. For our analysis we use the merger rates computed using the empirical Star Formation Rate Function (SFRF) as galaxy statistics and the Fundamental Metallicity Relation (FMR) to assign metallicity to galaxies (see Figure 8 of \citep{Boco:2020pgp}). In order to reduce the impact of the uncertainties in the astrophysical modeling, we decide to re-scale all the merger rates per unit comoving volume to match the local values measured by LIGO/Virgo/KAGRA \citep{LIGOScientific:2020kqk,LIGOScientific:2021qlt}: $23.9\substack{+14.9 \\ -8.6}\,\rm Gpc^{-3}\,yr^{-1}$ for BH-BH,  $320\substack{+490 \\ -240}\,\rm Gpc^{-3}\,yr^{-1}$ for NS-NS and $45\substack{+75 \\ -33}\,\rm Gpc^{-3}\,yr^{-1}$ for BH-NS.

\subsection{Cross-correlation technique}
Because the kernels of both our cosmic fields are broad functions of redshift, the angular cross-correlation can be computed using the Limber approximation \citep{Limber:1954zz} in the following way:

\begin{equation}
    C_{\ell}^{\kappa \Omega} = \int _{0}^{z_{\star}} \dfrac{dz}{c} \dfrac{H(z)}{\chi^2(z)}\,  W^{\kappa} (z) W^{\Omega}(z) \, P\biggl(k = \frac{l}{\chi(z)} ,z \biggl),
\end{equation}
where $P (k, z)$ is the matter power spectrum, which we computed using the \texttt{CLASS}\footnote{The Cosmic Linear Anisotropy Solving System, available at \url{http://class-code.net}} public code \citep{lesgourgues2011cosmic,Blas_2011}. The nonlinear evolution of the matter power spectrum was taken into account using the \texttt{HALOFIT} prescription \citep{Smith:2002dz}. 
Assuming that both the SGWB and the CMB lensing behave as Gaussian random fields, the variance of $C_{\ell}^{\kappa \Omega}$ is given by: 

\begin{equation}
    \bigl(\Delta C_{\ell}^{\kappa \Omega} \bigl)^{2} = \dfrac{1}{(2\ell + 1) f_{\rm{sky}}} \Bigl[ \bigl(C_{\ell}^{\kappa \Omega}\bigl)^{2} + \bigl(C_{\ell}^{\kappa \kappa} + N_{\ell}^{\kappa \kappa} \bigl)\bigl(C_{\ell}^{\Omega \Omega} \boldsymbol{+S_{\ell}^{\Omega \Omega}} + N_{\ell}^{\Omega \Omega} \bigl) \Bigl], 
\end{equation}
where $f_{\rm{sky}}$ is the sky fraction covered by both the SGWB and the CMB lensing surveys, $N_{\ell}^{\kappa \kappa}$ is the lensing noise, $S_{\ell}^{\Omega \Omega}$ and $N_{\ell}^{\Omega \Omega}$ are the SGWB the shot and instrumental noise respectively. For our analysis, we employ the Simons Observatory lensing noise curves \citep{Ade2019} and we adopt the fiducial Large Aperture Telescope value $f^{\kappa}_{\rm{sky}} = 0.4$ for the sky fraction. Since GW experiments cover the entire sky (i.e. $f^{\Omega}_{\rm{sky}} = 1$), we use the limiting value $f_{\rm{sky}} = f^{\kappa}_{\rm{sky}}$ for the cross-correlation. As for the SGWB, we compute the instrumental noise curves for different detector network configurations using the software \texttt{schNell}\footnote{The package is public available at \url{https://github.com/damonge/schNell}.} \citep{Alonso:2020rar}, while we evaluate the SGWB shot noise contribution through the novel map-making technique developed in \citep{Capurri:2021zli} and summarized in Appendix \ref{app:shot_noise}. The signal-to-noise ratio (S/N) at multipole $\ell$ is then given by:

\begin{equation} \label{eq:snr}
    \biggl( \dfrac{S}{N} \biggl)_{\ell}^{2} = \dfrac{\bigl( C_{\ell}^{\kappa \Omega} \bigl)^{2}}{\bigl(\Delta C_{\ell}^{\kappa \Omega} \bigl)^{2}} = \dfrac{(2\ell + 1) f_{\rm{sky}}\bigl( C_{\ell}^{\kappa \Omega} \bigl)^{2}}{\bigl(C_{\ell}^{\kappa \Omega}\bigl)^{2} + \bigl(C_{\ell}^{\kappa \kappa} + N_{\ell}^{\kappa \kappa} \bigl)\bigl(C_{\ell}^{\Omega \Omega} \boldsymbol{ + S_{\ell}^{\Omega \Omega}} + N_{\ell}^{\Omega \Omega} \bigl)}
\end{equation}
and the cumulative S/N for multipoles up to $\ell_{\rm{max}}$ is 

\begin{equation} \label{eq:cumulative_snr}
    \biggl( \dfrac{S}{N} \biggl) (\ell < \ell_{\rm{max}}) = \sqrt{\sum_{\ell = \ell_{\rm{min}}}^{\ell_{\rm{max}}} \biggl( \dfrac{S}{N} \biggl)_{\ell}^{2}} .
\end{equation}
Similarly, the auto-correlation power spectra can be evaluated as:

\begin{equation}
   C_{\ell}^{XX}  = \int _{0}^{z_{\star}} \dfrac{dz}{c} \dfrac{H(z)}{\chi^2(z)}  \bigl[W^{X} (z) \bigl]^{2} \, P\biggl(k = \frac{l}{\chi(z)} ,z \biggl),   
\end{equation}
where $X$ is either $\kappa$ or $\Omega$, and the associated S/N is given by: 

\begin{equation} \label{eq:snr_auto}
    \biggl( \dfrac{S}{N} \biggl)_{\ell}^{2} = \dfrac{(2\ell + 1)}{2} f^{X}_{\rm{sky}}  \dfrac{ \bigl( C_{\ell}^{XX} \bigl)^{2}}{\bigl( C_{\ell}^{XX} \boldsymbol{ + S_{\ell}^{XX}} + N_{\ell}^{XX} \bigl)^{2}}.
\end{equation}
where $S_{\ell}^{XX}$, $N_{\ell}^{XX}$ and $f^{X}_{\rm{sky}}$ are the shot noise, the instrumental noise and the sky fraction covered by the considered survey for the tracer X.

\section{Results} \label{sec:results}

\begin{specialtable} 
\caption{Location coordinates and orientation angles for the detectors considered in this work. For the future 3G detectors ET and CE, we adopted arbitrary locations and orientations, assuming that the detectors will be built somewhere in one of their proposed sites: Sardinia (Italy) and Utah (USA), respectively. See \citep{Alonso:2020rar} for further details on the definition of the coordinates and the orientation angles. \label{tab1}}
\begin{tabular}{cccc}
\toprule
\textbf{Detector}	& \textbf{Latitude} (deg)	& \textbf{Longitude} (deg) & \textbf{Orientation} (deg)\\
\midrule
LIGO Hanford		& 46.6			& -119.4  & 171.8\\
LIGO Livingston	   & 30.7			& -90.8  & 243.0\\
Virgo	   & 43.6			& 10.5  & 116.5\\
KAGRA	   & 36.3			& 137.2  & 225.0\\
ET$^*$	   & 40.1			& 9.0  & 90.0\\
CE$^*$	   & 40.8			& -113.8  & 90.0\\
\bottomrule
\end{tabular}
\end{specialtable}

We study the cross-correlation between the astrophysical SGWB and the CMB lensing convergence in two different scenarios. First, we examine the case where all GW events are taken into account when computing the SGWB energy density. Second, we consider the residual SGWB as potentially measured by a specific detector, i.e. we only take into account the contribution of the unresolved events. The events that can be resolved individually are filtered out by means of a signal-to-noise threshold, which we fix at the reference value of $\bar{\rho} = 8$ (see Appendix \ref{app:sgwb_characterization} for more details). We include in our analysis the present and future ground-based GW detectors LIGO/Virgo/KAGRA, ET and CE. Indeed, to detect a diffuse GW background, it is necessary to cross-correlate the outputs of several interferometers handled as a network. In Table \ref{tab1}, we report the detector coordinates and orientation angles that we plug in
the \texttt{schNell} to compute the noise curves for specific network configurations. In particular, for the anisotropies of the residual SGWB detected by LIGO/Virgo/KAGRA, we consider a network composed of the first four interferometers in Table \ref{tab1} and the associated noise curve. Instead, to investigate the detectability of the anisotropies of the residual SGWB measured by ET or CE, we consider a network composed of LIGO/Virgo/KAGRA and ET or CE, respectively. In any case, we consider an integration time of $T = 1$ yr. Finally, we also consider an extended network given by all the detectors taken into account in this work. We use its noise curves for $T = 1$ yr and $T = 10$ yrs to study the detectability of the total background, given by all resolved and unresolved events.  We work at the reference frequency of 65 Hz, which falls into the middle of the sensitivity bands of all the considered detectors.

\begin{figure} [t]
\centering
\includegraphics[width= 13 cm]{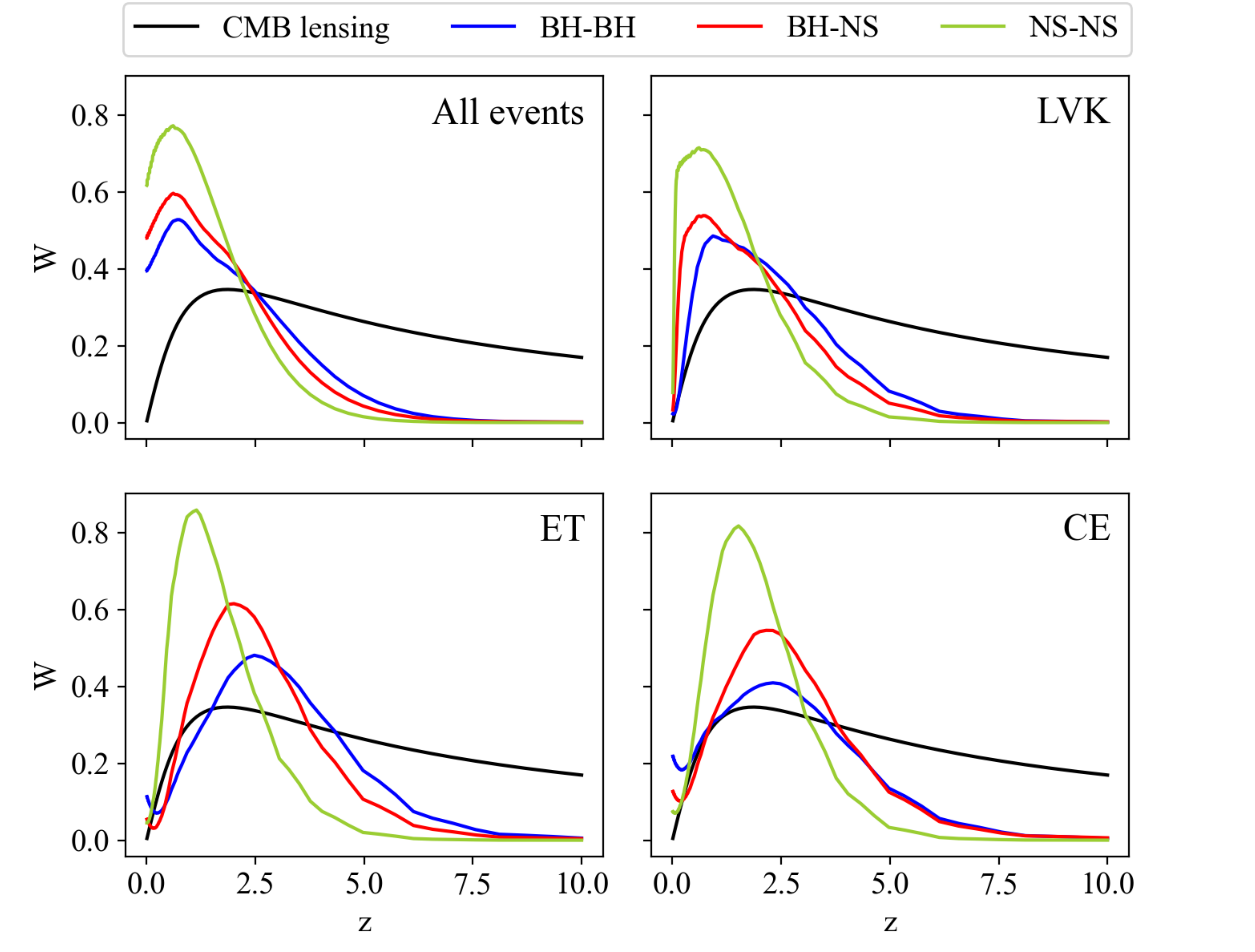}
\caption{SGWB kernel $W^{\Omega} (z)$ as a function of redshift for the three types of sources (BH-BH, BH-NS and NS-NS in blue, red and green respectively) at 65 Hz, compared with the lensing kernel $W^{\kappa}$ (in black). In the upper left panel we consider the SGWB given by the superposition of all GW events, resolved and unresolved. In the other three panels, instead, we consider the residual background obtained removing the GW events that are resolved by the considered detector (LIGO/Virgo/KAGRA, ET and CE respectively).  \label{fig:kernels}}
\end{figure}

In Figure \ref{fig:kernels}, we compare the CMB lensing kernel with the SGWB kernel for different sources (BH-BH, BH-NS and NS-NS) and detectors, computed using equations \ref{eq:lensing_kernel} and \ref{eq:sgwb_kernel}, respectively. 
The lensing kernel is a broad function of redshift that peaks at $z \simeq 2$ and decreases slowly at higher redshifts, making the CMB lensing convergence a powerful probe of LSS up to the last-scattering surface.
In the upper-left panel, we consider the SGWB obtained integrating all the GW events, resolved and unresolved. The kernel is quite broad and peaks at $z \lesssim$ 1 for all types of sources. The exact position of the peak and the shape of the kernel are the outcomes of the complex interplay among the behaviour of three factors: the merger rate, the flux carried by each GW event and the SGWB bias. On the one hand, the shape of the redshift distribution $d\Omega/ dz$ reflects the fact that the merger rate peaks with the cosmic star formation rate at $z \simeq 2$ and rapidly declines at higher redshifts. On the other hand, nearer events largely contribute to the SGWB energy density because their flux is less diluted. Finally, the SGWB bias is an increasing function of redshift (see upper panels of Figure \ref{fig:omega_tracer}) so that its effect is to give more weight to more distant objects. The BH-BH kernel is slightly broader than BH-NS and NS-NS ones because BH-BH binaries have a greater chirp mass and produce more energetic GW signals, contributing significantly up to higher redshift. The upper-right panel shows the kernels of the residual SGWB, which we obtain filtering out the events individually resolved by LIGO/Virgo/KAGRA. Typically, this operation removes many of the nearby events, especially the energetic BH-BH mergers, conferring a proportional higher weight to more distant ones. For this reason, the kernels are broader and peak at slightly higher redshifts than in the previous case, where we integrated all GW events to obtain the SGWB.

As we can see in the lower panels of Figure \ref{fig:kernels}, these effects are even more evident for 3G detectors such as ET and CE, which will be able to resolve BH-BH mergers individually up to $z \simeq 10$. As a result, the kernels are substantially broader than previous cases and peak at higher redshifts. It is also evident that these SGWB configurations present a more extended overlap with the CMB lensing kernel. As we will see in the following, this has a considerable impact on the cross-correlation results. The non-monotonic behaviour that characterizes the SGWB kernels at low redshifts for ET and CE is due to the lensing term in equation \ref{eq:sgwb_kernel}. As pointed out in Appendix \ref{app:sgwb_characterization} and extensively discussed in \citep{Capurri:2021zli}, the effect of lensing is to reduce the SGWB energy density, boosting some events above the detection threshold and making them directly resolvable by the instrument so that they do not contribute to the SGWB. This process is more effective for 3G detectors and produces a dip at $z\simeq 0.2-0.3$. 

\begin{figure} [t]
\centering
\includegraphics[width= 13 cm]{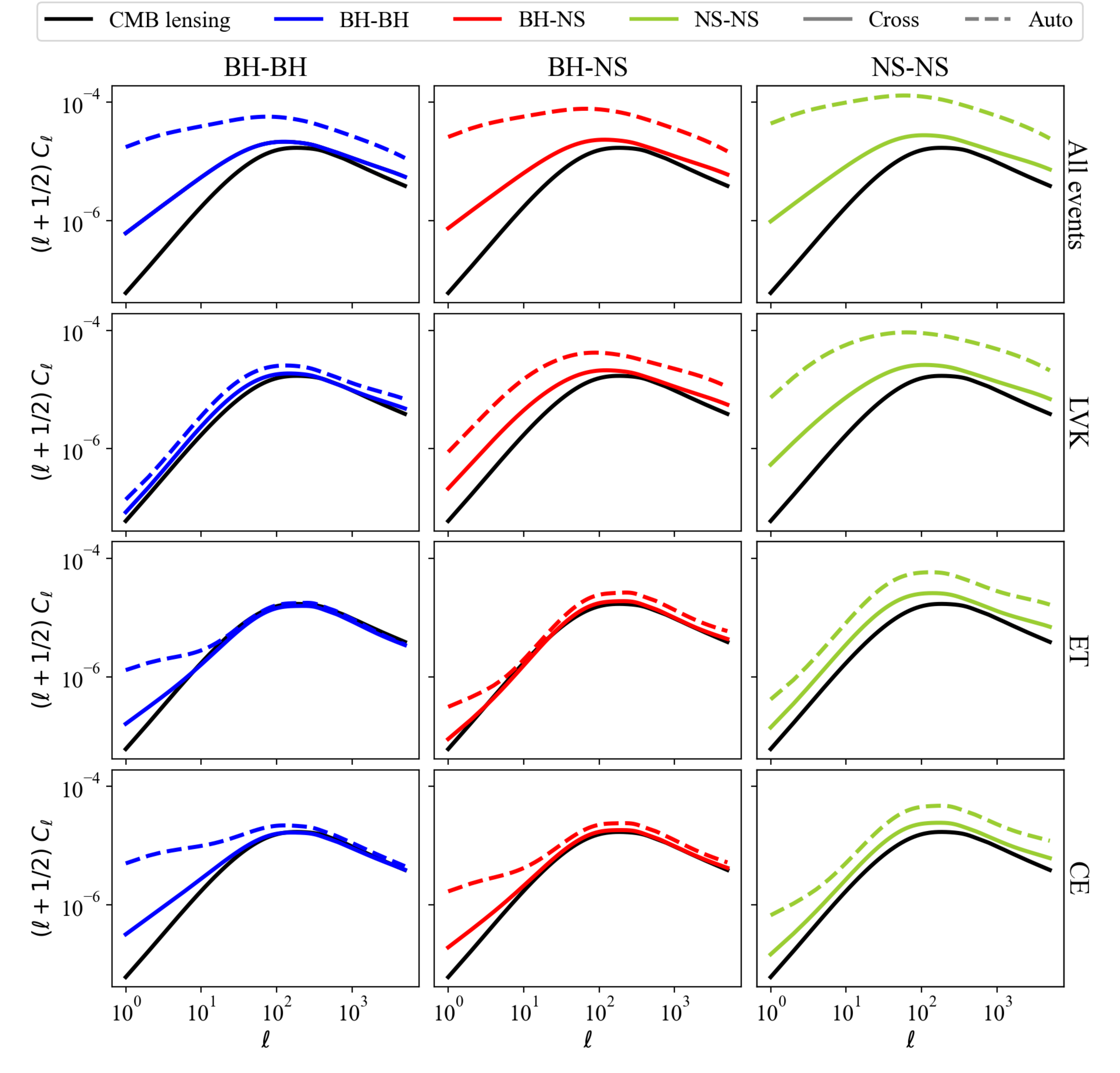}
\caption{Auto- and cross-correlation angular power spectra for SGWB and CMB lensing convergence anisotropies. Each panel displays the SGWB auto-correlation angular power spectrum (dashed) and the SGWB$\times$CMB convergence cross-correlation angular power spectrum (solid) for given source (BH-BH, BH-NS or NS-NS) and detector (LIGO/Virgo/KAGRA, ET, CE or no detector, i.e. all events considered) and the convergence auto-correlation angular power spectrum for reference. \label{fig:cls}}
\end{figure}

In Figure \ref{fig:cls}, we show the auto- and cross-correlation power spectra for all the sources and detectors. Although we perform all the calculations in the Limber approximation, the SGWB auto-correlation power spectra are in good agreement\footnote{In this work we opt for a simpler analytic treatment with respect to the one adopted in \citep{Capurri:2021zli}, neglecting some of the relativistic effect that are instead taken into account using \texttt{CLASS} to compute the power spectra. However, we verified that the results obtained with the two pipelines are compatible within a factor of order unity, which do not affect our overall results.} with the results obtained in \citep{Capurri:2021zli} with the public code \texttt{CLASS} \citep{lesgourgues2011cosmic,Blas_2011}. 
At large angular scales, all the spectra behave as a power-law whose slope depends on the source type and the detector. For ET and CE, the large-scale behavior is significantly affected by the magnification bias and the power-law is broken. 
As expected, the $\Omega \times \kappa$ cross-correlation power spectrum always lies between the two auto-correlation power spectra. The fact that the cross-correlation is strong is quite remarkable and constitutes one of the most important findings of this paper since it proves that the large-scale distribution of late-time objects such as merging compact binaries is well correlated with the linear structures probed by CMB lensing. 

\begin{figure} [t]
\centering
\includegraphics[width= 13 cm]{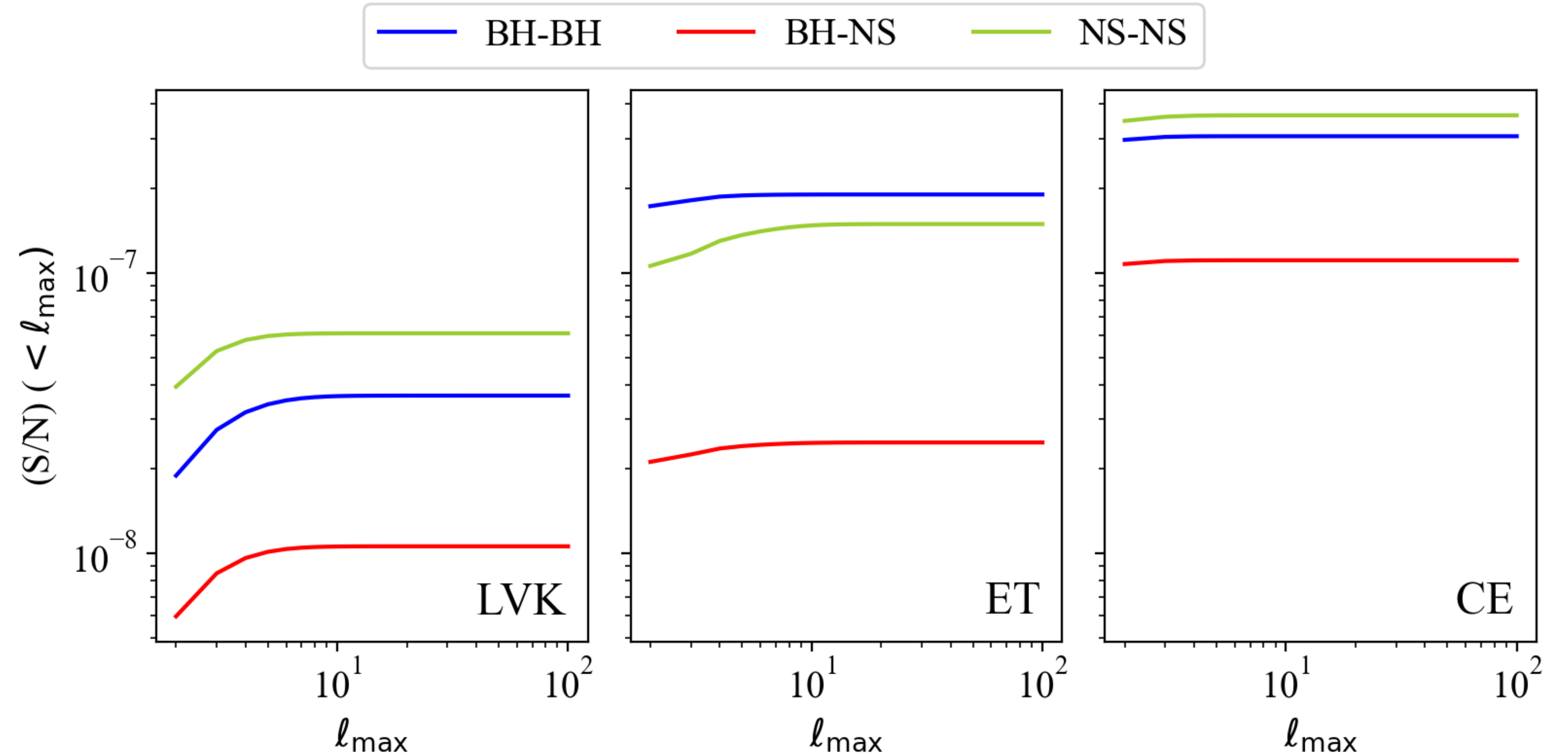}
\caption{Cumulative S/N as a function of $\ell_{\rm{max}}$ for the SGWB auto-correlation angular power spectrum. The curves have been evaluated summing the  equation \ref{fig:snr_auto} from $\ell_{\rm{min}} = 1$ up to $\ell_{\rm{max}}$. \label{fig:snr_auto}}
\end{figure}

The subtle SGWB anisotropies produced by the LSS, described by the power spectra in Figure \ref{fig:cls}, coexist with the much larger ones caused by the spatial and temporal discreteness of the GW sources that make up the SGWB. 
Indeed, the shot noise power spectrum outreaches the SGWB auto-correlation power spectrum by several orders of
magnitude (see e.g. Figure \ref{fig:shot_noise} and related comments in Appendix \ref{app:shot_noise}). Moreover, the instrumental noise is an even more significant killing factor for measuring intrinsic SGWB anisotropies. For all the considered detector networks, the noise curve $N_{\ell}$ is orders of magnitude larger than the shot noise power spectrum $S_{\ell}$. For these reasons, as we can see in Figure \ref{fig:snr_auto}, the S/N of the auto-correlation power spectra are broadly smaller than unity for all types of sources detector configurations.

\begin{figure} 
\centering
\includegraphics[width= 13 cm]{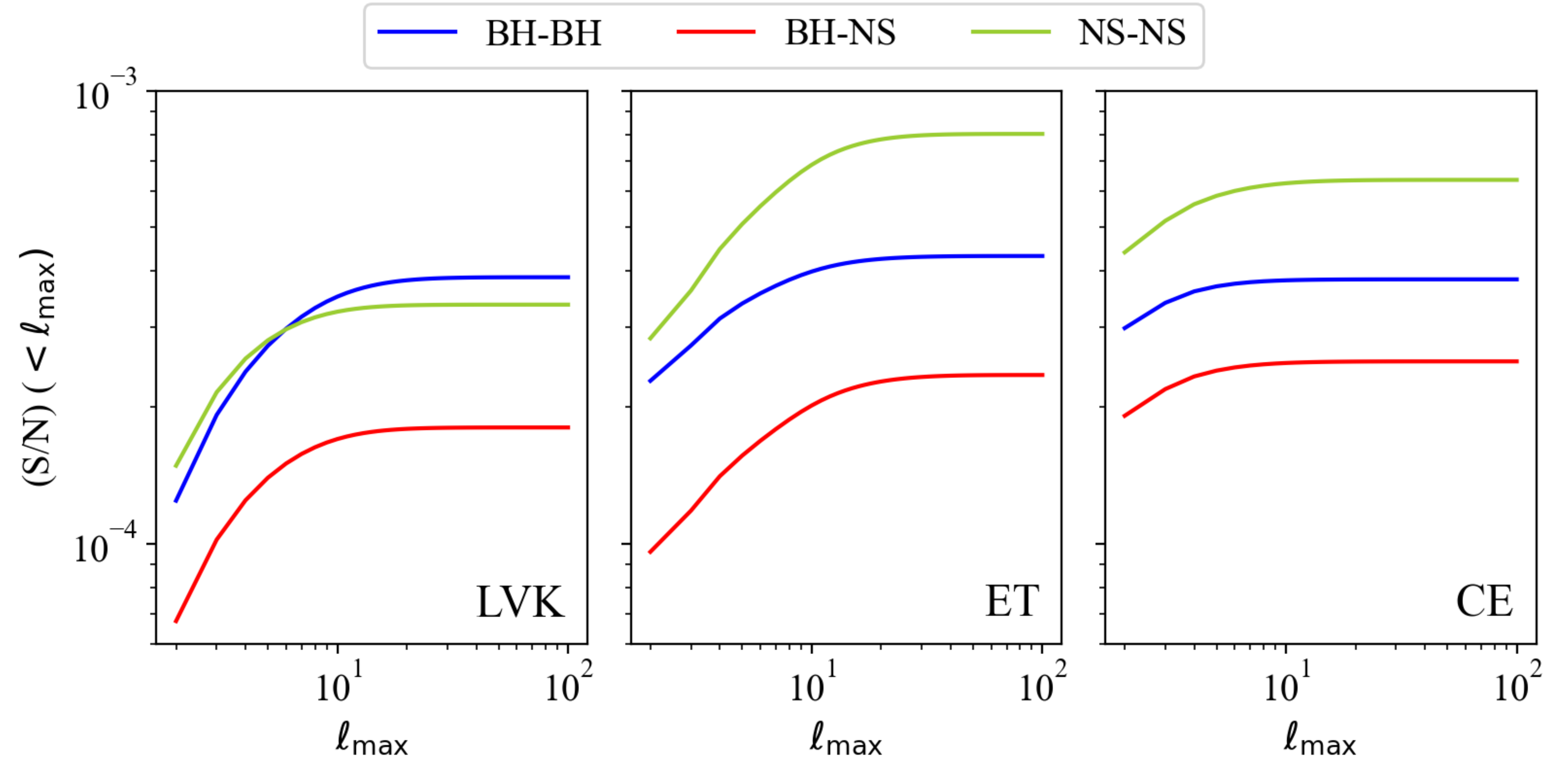}
\caption{Cumulative S/N for $\Omega \times \kappa$ cross-correlation angular power spectrum as a function of $\ell_{\rm{max}}$. The curves are evaluated by means of equations \ref{eq:snr} and \ref{eq:cumulative_snr} starting from $\ell_{\rm{min}} = 1$. The color code is the same as in Figure \ref{fig:kernels}. \label{fig:snr_cross}}
\end{figure}

The cumulative S/N of the $\Omega \times \kappa$ cross-correlation, displayed in Figure \ref{fig:snr_cross}, shows that cross-correlating with CMB lensing convergence is an effective way to mitigate the impact of instrumental and shot noise: indeed, it is around three orders of magnitude larger than the one of auto-correlation. The cross-correlation with another tracer of the same underlying dark matter distribution actually enhances the SGWB anisotropies induced by the LSS. Unfortunately, the instrumental noise is too high and this substantial improvement is not sufficient to guarantee a direct detection of the cross-correlation signal.

\begin{figure} [t]
\centering
\includegraphics[width= 12 cm]{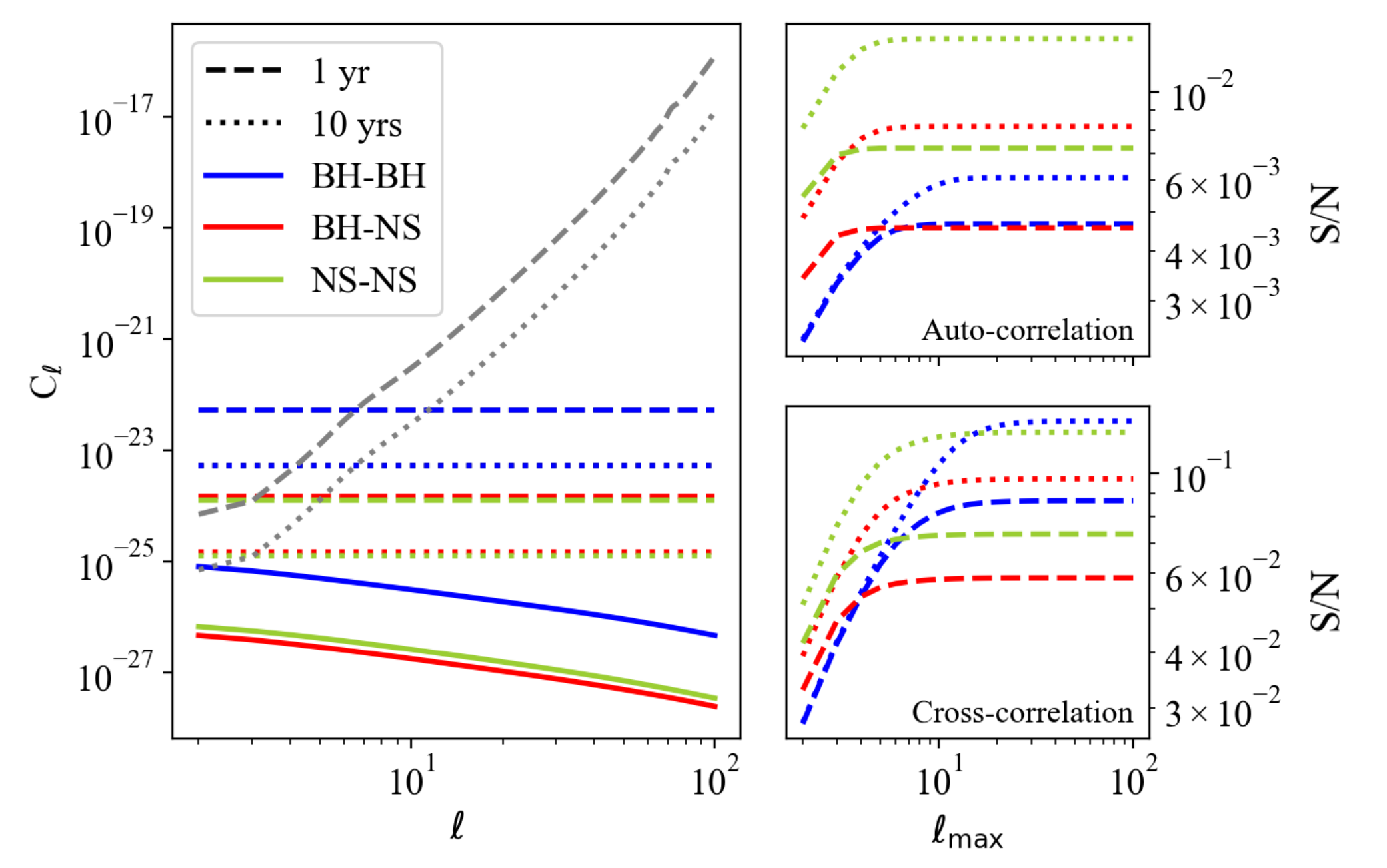}
\caption{Left panel: the grey lines represent noise curves $N_{\ell}$ for the extended network given by LIGO/Virgo/Kagra + ET + CE, for two different integration times:  $T = 1$ yr (dashed) and $T = 10$ yrs (dotted). The noise curves are compared with the shot noise power spectrum $S_{\ell}$ for the two different integration times and the auto-correlation power spectrum $C_{\ell}$ for the anisotropies of total SGWB given by all the resolved and unresolved events. Right panels: auto- and cross-correlation cumulative S/N. The different line styles represent different integration times, as specified in the left panel. The colour code is the same as in the previous figures.   \label{fig:lvkec}}
\end{figure}

As a final step, we try to increase as much as possible the S/N using a network with the five instruments considered in this work to detect the total SGWB given by all resolved and unresolved events. Combining the outputs of more detectors is an effective way to reduce the noise curve $N_{\ell}$, whereas considering the total background enhances the signal because the amplitude of the monopole is higher than the one of the residual background. Moreover, we also explore the benefits of measuring the GW signal for a longer integration time, $T = 10$ yrs. In Figure \ref{fig:lvkec} we show the results of this analysis. As we can see in the left panel, where we show the noise curves for $T = 1$ yr and $T = 10$ yrs together with the shot noise and auto-correlation power spectra, the improved sensitivity of the extended network allows to observe the flat shot noise power spectrum. Unfortunately, the amplitude of the intrinsic anisotropies is well below the noise level, as we can see in the right panels: despite a two-orders-of-magnitude enhancement, the cumulative S/N of both auto- and cross-correlation is still too low to allow a direct detection even for $T = 10$ yrs.

\section{Conclusion} \label{sec:conclusion}
The anisotropies of the SGWB produced by merging compact binaries in galaxies contain a wealth of information about the physical properties of GW emitters, their distribution in redshift and their position in the sky. However, the shot noise caused by the spatial and temporal discreteness of the GW emitters covers the signal. Moreover, the poor angular resolution of GW instruments constitutes a significant obstacle to the detection of SGWB anisotropies.

Other works have already shown that cross-correlating the SGWB with other cosmological probes mitigates the impact of shot noise. In this paper, we studied the potential of cross-correlating the SGWB with CMB lensing. We produced forecasts for the cross-correlation of forthcoming high precision measurements of the SGWB energy density and the CMB lensing convergence. Specifically, we considered LIGO/Virgo/KAGRA at design sensitivity, ET and CE for the gravitational waves and the SO for the CMB. 

Starting from a detailed model of stellar and galactic astrophysics, based on simulations and observationally driven prescriptions, we characterized the SGWB energy density as a tracer of LSS, derived its kernel and compared it with the CMB lensing one. 
We computed the auto- and cross-correlation power spectra in the Limber approximation for three types of SGWB sources: BH-BH, BH-NS and NS-NS merging binaries. The cross-correlation power spectrum always lies between the auto-correlation power spectra, showing a good correlation between the two cosmic fields for each source and detector. This result is not trivial since it implies that the distribution of merging compact binaries traced by the SGWB anisotropies is well correlated with the linear structures probed by CMB lensing.
Finally, we computed the S/N for both the auto- and the cross-correlation power spectra. To this purpose, we used the shot noise estimation naturally provided by a novel framework to distribute the GW emitters in the sky and computed the instrumental noise curves with the public package \texttt{schNell}. 

We found that the auto-correlation cumulative S/N is extremely low, since the instrumental noise outreaches both the signal and the shot noise by several orders of magnitude, even when considering third-generation GW detectors. The cross-correlation with CMB lensing effectively enhances the S/N of at least three orders of magnitude, but this improvement is not sufficient to ensure a direct detection of the intrinsic SGWB anisotropies induced by the LSS.  
On the other hand, we found that by combining all the instruments considered in this work as a single network operating for ten years, it will be possible at least to constrain the shot noise contribution.

In conclusion, we have shown that cross-correlating the astrophysical SGWB with CMB lensing improves the S/N by orders of magnitude. However, we found that the instrumental noise constitutes a major limiting factor for the S/N of both auto- and cross-correlation. Nevertheless, this study represents a novel aspect, directly impacting the characterization of the SGWB anisotropies produced by the LSS.

\vspace{6pt}

\authorcontributions{All authors have read and agreed to the published version of the manuscript.}

\funding{AL acknowledges funding from the EU H2020-MSCA-ITN-2019 Project 860744 \textit{BiD4BESt: Big Data applications for black hole Evolution STudies} and from the PRIN MIUR 2017 prot. 20173ML3WW, \textit{Opening the ALMA window on the cosmic evolution of gas, stars and supermassive black holes}. GC and CB are partially supported by the INDARK INFN grant. CB acknowledges support from the COSMOS $\&$ LiteBIRD Networks by the Italian Space Agency (\url{http://cosmosnet.it})}

\acknowledgments{We warmly thank L. Boco, E. Komatsu,  D. Spergel, T. Ronconi and A. Lonappan for useful discussions. We thank the two anonymous referees for their careful reading of the manuscript and their insightful suggestions, which helped improve the robustness and the interpretation of our results.}

\conflictsofinterest{The authors declare no conflict of interest.}

\appendixtitles{no} 
\appendixstart
\appendix

\section{} \label{app:sgwb_characterization}
\begin{figure} 
\centering
\includegraphics[width= 10 cm]{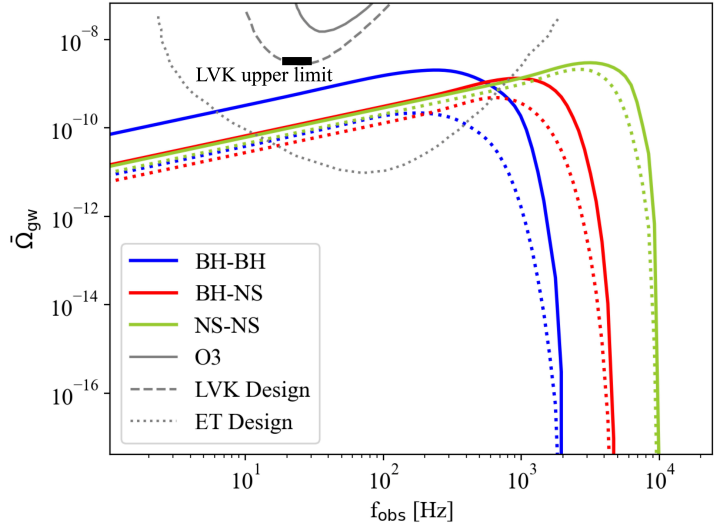}
\caption{ Isotropic SGWB energy density as a function of frequency for BH-BH (blue), BH-NS (red) and NS-NS (green). The solid curves represent the total background, given by the superposition of all the GW events, resolved and unresolved. The dotted curves, instead, represent the residual background for ET, obtained filtering out those events whose signal-to-noise ratio $\rho$ is high enough to be resolved by the detector. The grey curves are the power-law integrated sensitivity curves \citep{Thrane:2013oya} for ET and LIGO/Virgo/KAGRA O3 and design sensitivity, as specified in the legend. We also show the current upper limit on the isotropic SGWB: $\omegagw \leq 3.4 \times 10^{-9}$ at 25 Hz for a power-law background with a spectral index of 2/3, expected for compact binary coalescences \citep{KAGRA:2021kbb}.} \label{fig:isotropic_omega}
\end{figure}

In this Appendix we briefly review the theoretical modeling of the SGWB developed in \citep{Capurri:2021zli} and adopted throughout this work. The main ingredient for the description of the SGWB is its isotropic energy density $\baromega(\fobs)$ at the observed frequency $\fobs$. This quantity can be computed summing the contributions of all the GW events whose signal-to-noise is below a given detection threshold:
\begin{equation} \label{eq:omega_gw}
\baromega(\fobs)= \dfrac{8 \pi G \fobs} {3 H_{0}^{3} c^{2}} \int dz \int d \mc \;
\dfrac{R_{\rm{merge}}(\mc,z)}{(1+z)\; h(z) } \dfrac{dE}{df} (f_{e}(z) | \mc) \int_{0}^{\infty}  d \rho \, \epsilon_{\bar{\rho}}(\rho)\, P_{\rho} (\rho | \mc, z) ,
\end{equation}
where $\mc$ is the chirp mass, $f_{e} = (1+z) \fobs$ it the source frequency, $\rho$ is the signal-to-noise ratio associated to a certain GW event and $\bar{\rho}$ is the detection threshold.
In the previous expression, $R_{\rm{merge}}= d^{2} \dot{N} / dV d\mc$ is the intrinsic merger rate per unit comoving volume and per unit chirp mass, which we compute following the prescriptions described in  \citep{Boco:2019teq, Boco:2020pgp}, $h(z) = [ \Omega_{M}(1+z)^{3} + 1 - \Omega_{M}]^{1/2}$ accounts for the dependence of comoving volume on cosmology, $dE/df (f_{e}(z) | \mc)$ is the energy spectrum of the signal emitted by a single binary \citep{Ajith:2007kx}, $P_{\rho} (\rho | \mc, z)$ is the sky-averaged distribution of signal-to-noise ratio for a given detector at given chirp mass and redshift \citep{Finn:1995ah,Taylor:2012db} and $\epsilon_{\bar{\rho}} (\rho)$ is an efficiency function that set the sharpness of the signal-to-noise threshold. In the simplest case, $\epsilon$ is a step function centered on $\bar{\rho}$, so that its effect is to sharply remove all the GW events whose signal-to-noise ratio is greater than $\bar{\rho}$. In this work we opt for a smooth threshold given by an error function with amplitude equal to the square root of the variance of the sky-averaged distribution of the signal-to-noise ratio. Of course, the energy density of the SGWB produced by all the events, resolved and unresolved, can be easily obtained using $\bar{\rho} = \infty$ in equation \ref{eq:omega_gw}: in this way, the last integral is equal to 1 and all the events are taken into account. In figure \ref{fig:isotropic_omega} we show the isotropic energy density of the total SGWB given by all the events, resolved and unresolved, and the residual SGWB measured by ET.

In order to characterize the SGWB as a tracer of matter and evaluate its kernel by means of equation \ref{eq:sgwb_kernel}, we need to define its redshift distribution $d\Omega / dz$, its bias $b_{\Omega}$ and its magnification bias $s_{\Omega}$. The redshift distribution is easily obtained from equation \ref{eq:omega_gw} removing the integral in redshift space: in this way, we obtain the contribution to the energy density coming from the GW events at redshift $z$.

The bias $b_{\Omega}$ quantifies the mismatch between the distribution of the SGWB energy density and the underlying dark matter density. Since the SGWB traces the distribution of the galaxies that host the compact binaries, the bias $b_{\Omega}$ is directly related to the galaxy bias. Following the approach of \citep{Boco:2019teq,Boco:2020pgp}, we classify the host galaxies by means of their star formation rate $\psi$ and we define the bias $b_{\Omega}$ as the following weighted mean of the galaxy bias $b(z,\psi)$: 

\begin{equation} \label{eq:b_omega}
    b_{\Omega}(z,\fobs) 
    = \dfrac{\int d \log_{10}\psi \dfrac{d^{2} \baromega }{dz d \log_{10}\psi}(z,\psi,\fobs) \; b(z,\psi)}
    {\int d \log_{10}\psi \dfrac{d^{2} \baromega}{dz d \log_{10}\psi}(z,\psi,\fobs)}\, ,
\end{equation}
where the weight is the SGWB energy density per unit redshift and star formation rate, which can be obtained from equation \ref{eq:omega_gw} using the merger rate per unit volume, chirp mass and star formation rate, $d^{3} \dot{N} / d\mc \, dV \, d \log_{10}\psi$, instead of the merger rate per unit volume and chirp mass. The galaxy bias as a function of the star formation rate $b(z,\psi)$ is computed from the bias as a function of the dark matter halo mass $b(z,M_{H})$ by beans of the abundance matching technique described in \citep{Aversa:2015bya}. The bias $b(z,M_{H})$, in turn, is computed as in \citep{Sheth:1999su} and approximated as in \citep{Lapi:2014ija}. 

The magnification bias $s_{\Omega}$ describes how the weak lensing affects the SGWB energy density. Adapting the general derivation that is found in the appendix of reference \citep{Hui_2007} to the case of SGWB, we define the magnification bias of the SGWB energy density through
\begin{equation}
    \dfrac{d\baromega^{\rm{lensed}}}{dz} \equiv \dfrac{d\baromega}{dz} \bigl[1+ \kappa(s_{\Omega} -1)], 
\end{equation}
where $\kappa$ is the lensing convergence. After some manipulation, we obtain the explicit expression 
\begin{equation} \label{eq:s_omega}
    s_{\Omega, \bar{\rho}}(z,\fobs) = - \dfrac{1}{2} \dfrac{d \log_{10} \bigl( \frac{d \baromega (\fobs,z, <\rho)}{dz}\bigl)}{d \log_{10} \rho} \Biggl| _{\rho = \bar{\rho}}\, ,
\end{equation}
which is similar to the one derived in \citep{Scelfo:2018sny,Scelfo:2020jyw} for resolved GW events: a crucial difference is that we are now considering only the GW events that are below the detection threshold $\bar{\rho}$.

\begin{figure} 
\centering
\includegraphics[width= 13 cm]{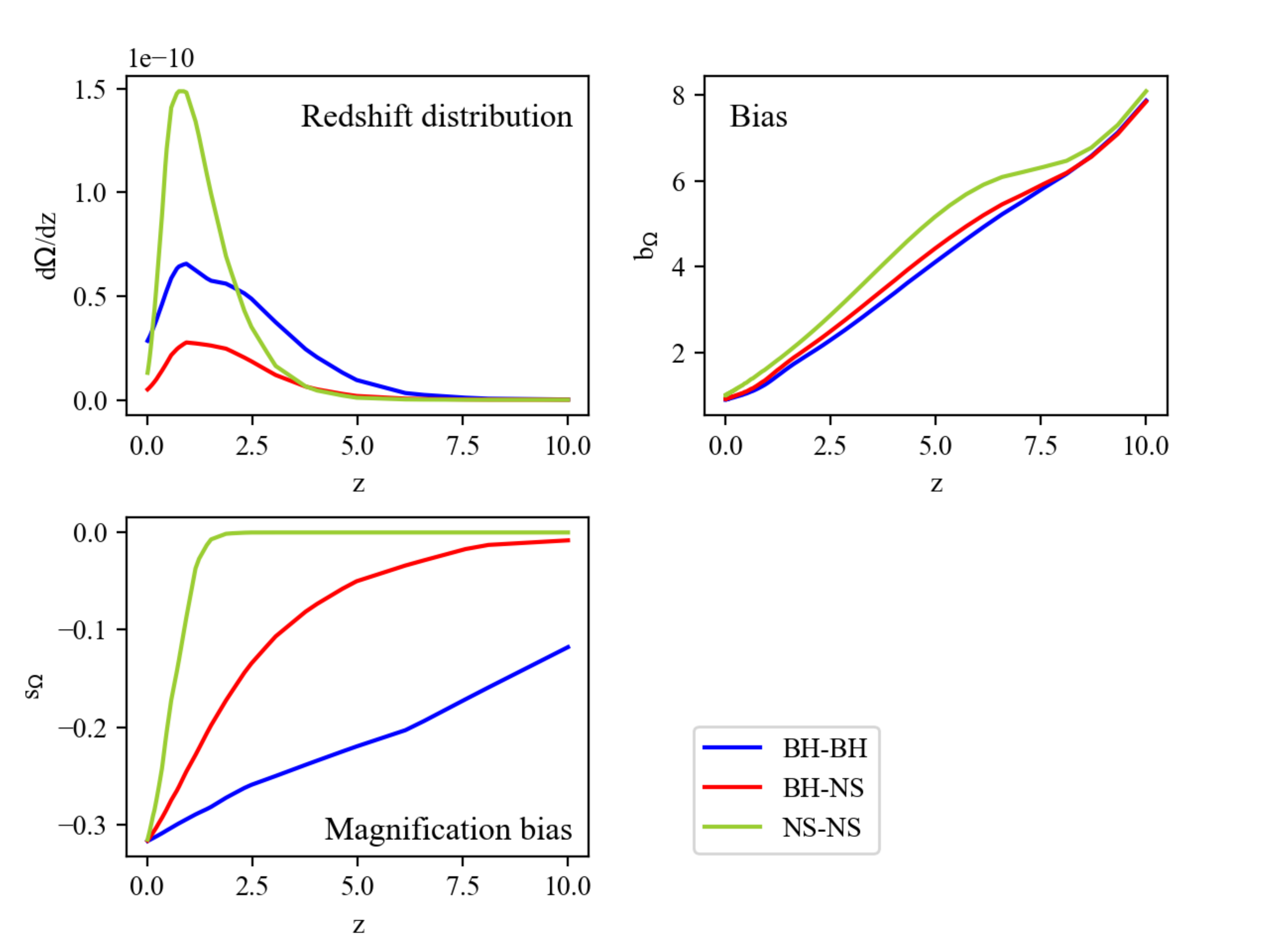}
\caption{Redshift distribution, bias and magnification bias of the SGWB energy density, evaluated for BH-BH (blue), BH-NS (red) and NS-NS (green) for ET at 65 Hz.} \label{fig:omega_tracer}
\end{figure}

For illustrative purposes, in Figure \ref{fig:omega_tracer} we show the redshift distribution, the bias and the magnification bias of the SGWB energy density for ET at 65 Hz.

\section{} \label{app:shot_noise}

One of the crucial points of our analysis is the estimation of the shot noise contribution to the SGWB anisotropies power spectrum. To this purpose, we rely on a novel framework to distribute the GW emitters in the sky and simulate a full-sky map of the expected signal. In this Appendix we summarize the main steps of this technique, referring the interested readers to  \citep{Capurri:2021zli} for more details.

First of all, we randomly distribute the GW emitters in the sky, according to the the mean number of unresolved events per unit time in each pixel:
\begin{equation}
    \langle \dot{N} \rangle = \dfrac{4 \pi}{N_{\rm{pix}}} \int dz \int d \mc \dfrac{d^{3} \dot{N}}{d \omega \, dz \, d\mc} \int_{0}^{\infty}  d \rho \, \epsilon_{\bar{\rho}}(\rho)\, P_{\rho} (\rho | \mc, z). 
\end{equation}

\begin{equation}
    \langle \dot{N} \rangle = \dfrac{4 \pi}{N_{\rm{pix}}} \int dz \int d \mc \dfrac{d^{3} \dot{N}}{d \omega \, dz \, d\mc} \int_{0}^{\bar{\rho}}  d \rho \, P_{\rho} (\rho | \mc, z). 
\end{equation}

In the previous expression, $N_{\rm{pix}}$ is the number of pixels in the map and $d \dot{N}/d \omega \, dz \, d\mc$ is the merger rate per unit redshift, solid angle and chirp mass, which can be easily obtained from the differential merger rate per unit chirp mass and comoving volume through: 
\begin{equation}
     \dfrac{d^{3} \dot{N}}{d \omega dz d\mc} = \dfrac{d^{2} \dot{N}}{dV d\mc}  \dfrac{c \, r(z)}{H_{0} \, h(z)}\,,
\end{equation}
where $r(z)$ is the comoving distance. 

At this point, we assign to each pixel a number of events occurring during a fixed observation time $T$, extracting it from a Poisson distribution with mean $\langle \dot{N}_{\rm{pix}} \rangle \times T $. We then assign to each event in each pixel a chirp mass and a redshift that we generate randomly from a 2D probability distribution obtained from the differential merger rate $d \dot{N}/ dz \, d \mc$. In this way, we can compute the energy density in each pixel summing the contributions from all the events:
\begin{equation}
    \Omega^{\rm{Poiss}}_{\rm{gw}} = \dfrac{8 \pi G \fobs}{3 H_{0}^{2} c^{3}} \, \dfrac{1}{T} \, \sum_{i=1}^{ \langle \dot{N} \rangle} \, \dfrac{\frac{dE}{df} (z_{i}, \mathcal{M}_{c \, i})}{4 \pi (1+z_{i}) r^{2}(z)}\,.
\end{equation}
The associated energy density contrast is given by:

\begin{equation}
    \delta^{\rm{Poiss}}_{\rm{gw}} = \dfrac{\omegagw^{\rm{Poiss}} - \langle \omegagw \rangle}{ \langle \omegagw \rangle}\,.
\end{equation}
In this way, we obtained map of the SGWB energy density contrast that follows a pure Poisson statistics. In order to introduce the clustering induced by the LSS, first of all we use the HEALPix\footnote{http://healpix.sourceforge.net} package \citep{Zonca2019,2005ApJ...622..759G} to compute the harmonic coefficients $a_{\ell m}^{\rm{Poiss}}$ of the Poissonian energy density contrast map. Subsequently, we introduce the correlation described by the theoretical angular power spectrum $C_{\ell}^{\Omega \Omega}$ in the following way: 
\begin{equation}
    a_{\ell m} = a_{\ell m}^{\rm{Poiss}} \dfrac{\sqrt{C_{\ell}^{\rm{Poiss}} + C_{\ell}^{\Omega \Omega}}}{\sqrt{C_{\ell}^{\rm{Poiss}}}}\,.
\end{equation}
Finally, we perform an inverse harmonic transform in order to obtain a map of the energy density contrast $\delta_{\rm{gw}}$, which contains both the shot noise contribution and the intrinsic correlation induced by the LSS. The energy density at each pixel can be easily computed as $\omegagw= \langle \omegagw \rangle (1+ \delta_{\rm{gw}})$. 

\begin{figure} 
\centering
\includegraphics[width= 13 cm]{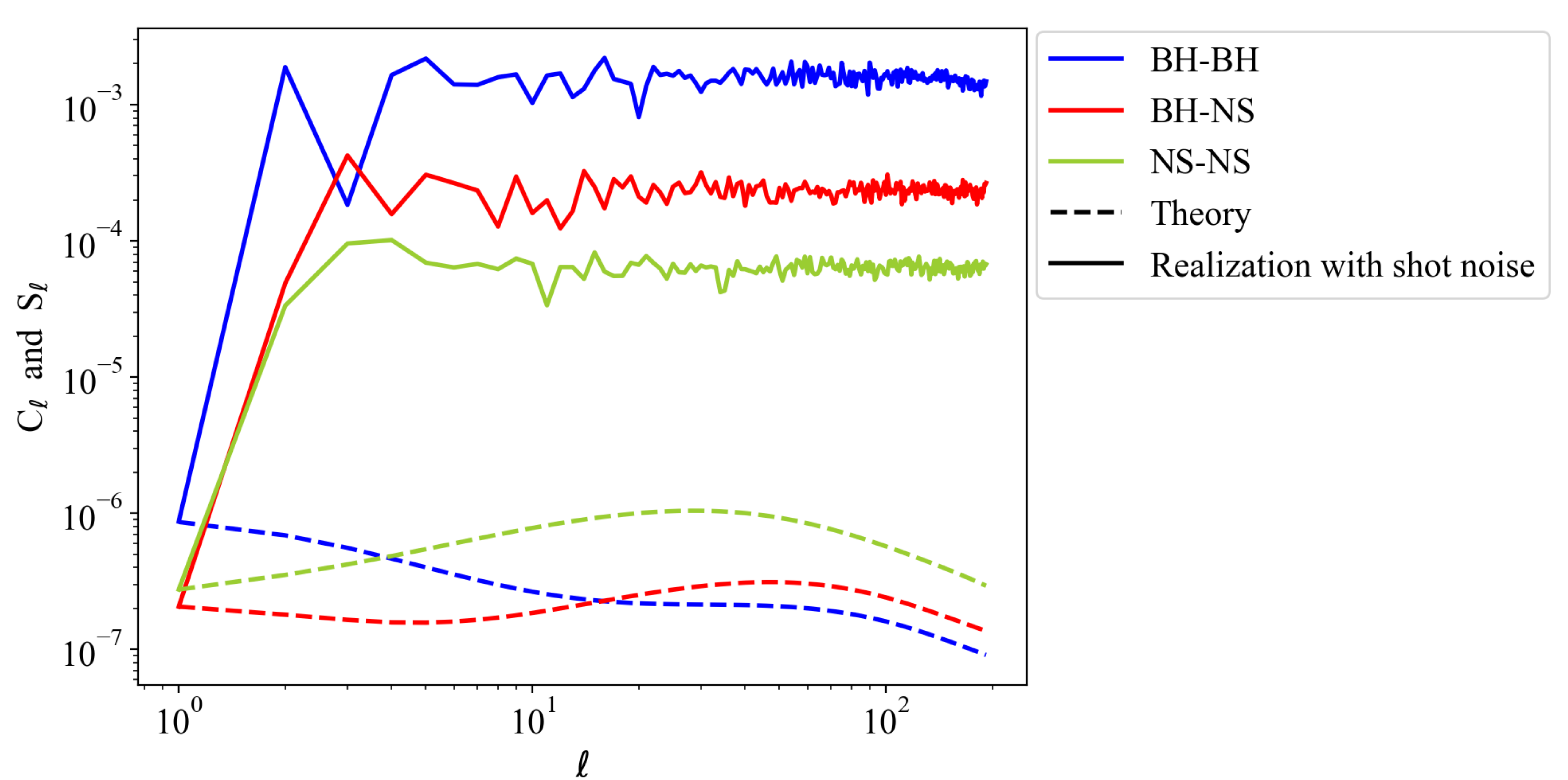}
\caption{Power spectrum of the SGWB anisotropies at 65 Hz, measured by ET during an observation time of T = 1 yr for BH-BH (blue), BH-NS (red) and NS-NS (green). The dashed curves are the theoretical auto-correlation power spectra shown in Figure \ref{fig:snr_auto}. The solid curves are the power spectra of specific realizations of the SGWB that contain also the intrinsic shot noise contribution, a flat offset that outreaches the intrinsic correlation by orders of magnitude.} \label{fig:shot_noise}
\end{figure} 

For illustrative purposes, in Figure \ref{fig:shot_noise} we show the angular power spectrum of three realizations of the SGWB (one for each type of source) at the reference frequency of 65 Hz, measured by ET during an observation of one year. For comparison, we also plot the theoretical power spectra that represent the correlation of the SGWB anisotropies induced by the LSS: the shot noise outreaches the signal by several orders of magnitude. The shot noise contribution is particularly high in the BH-BH case, because BH-BH mergers have a lower rate with respect to BH-NS and NS-NS ones. Of course, the amplitude of the shot noise is also determined by choice of the integration time T, as $S_{\ell} \propto T^{-1} $. In this work, we adopt a reference values $T = 1$ yr and $T = 10$ yrs and we estimate the shot noise amplitude averaging over 20 realizations of the SGWB.

\end{paracol}

\reftitle{References}

\bibliography{refs_sgwb_cmb}

\begin{thebibliography}{999}

\bibitem[Abbott \em{et~al.}(2016)Abbott et~al.]{LIGOScientific:2016aoc}
Abbott, B.P.; others.
\newblock {Observation of Gravitational Waves from a Binary Black Hole Merger}.
\newblock {\em Phys. Rev. Lett.} {\bf 2016}, {\em 116},~061102,
  \href{http://xxx.lanl.gov/abs/1602.03837}{{\normalfont
  [arXiv:gr-qc/1602.03837]}}.
\newblock
  doi:{\changeurlcolor{black}\href{https://doi.org/10.1103/PhysRevLett.116.061102}{\detokenize{10.1103/PhysRevLett.116.061102}}}.

\bibitem[Aasi \em{et~al.}(2015)Aasi et~al.]{LIGOScientific:2014pky}
Aasi, J.; others.
\newblock {Advanced LIGO}.
\newblock {\em Class. Quant. Grav.} {\bf 2015}, {\em 32},~074001,
  \href{http://xxx.lanl.gov/abs/1411.4547}{{\normalfont
  [arXiv:gr-qc/1411.4547]}}.
\newblock
  doi:{\changeurlcolor{black}\href{https://doi.org/10.1088/0264-9381/32/7/074001}{\detokenize{10.1088/0264-9381/32/7/074001}}}.

\bibitem[Acernese \em{et~al.}(2015)Acernese et~al.]{VIRGO:2014yos}
Acernese, F.; others.
\newblock {Advanced Virgo: a second-generation interferometric gravitational
  wave detector}.
\newblock {\em Class. Quant. Grav.} {\bf 2015}, {\em 32},~024001,
  \href{http://xxx.lanl.gov/abs/1408.3978}{{\normalfont
  [arXiv:gr-qc/1408.3978]}}.
\newblock
  doi:{\changeurlcolor{black}\href{https://doi.org/10.1088/0264-9381/32/2/024001}{\detokenize{10.1088/0264-9381/32/2/024001}}}.

\bibitem[Somiya(2012)]{Somiya:2011np}
Somiya, K.
\newblock {Detector configuration of KAGRA: The Japanese cryogenic
  gravitational-wave detector}.
\newblock {\em Class. Quant. Grav.} {\bf 2012}, {\em 29},~124007,
  \href{http://xxx.lanl.gov/abs/1111.7185}{{\normalfont
  [arXiv:gr-qc/1111.7185]}}.
\newblock
  doi:{\changeurlcolor{black}\href{https://doi.org/10.1088/0264-9381/29/12/124007}{\detokenize{10.1088/0264-9381/29/12/124007}}}.

\bibitem[Abbott \em{et~al.}(2019)Abbott et~al.]{LIGOScientific:2018mvr}
Abbott, B.P.; others.
\newblock {GWTC-1: A Gravitational-Wave Transient Catalog of Compact Binary
  Mergers Observed by LIGO and Virgo during the First and Second Observing
  Runs}.
\newblock {\em Phys. Rev. X} {\bf 2019}, {\em 9},~031040,
  \href{http://xxx.lanl.gov/abs/1811.12907}{{\normalfont
  [arXiv:astro-ph.HE/1811.12907]}}.
\newblock
  doi:{\changeurlcolor{black}\href{https://doi.org/10.1103/PhysRevX.9.031040}{\detokenize{10.1103/PhysRevX.9.031040}}}.

\bibitem[Abbott \em{et~al.}(2021{\natexlab{a}})Abbott
  et~al.]{LIGOScientific:2020ibl}
Abbott, R.; others.
\newblock {GWTC-2: Compact Binary Coalescences Observed by LIGO and Virgo
  During the First Half of the Third Observing Run}.
\newblock {\em Phys. Rev. X} {\bf 2021}, {\em 11},~021053,
  \href{http://xxx.lanl.gov/abs/2010.14527}{{\normalfont
  [arXiv:gr-qc/2010.14527]}}.
\newblock
  doi:{\changeurlcolor{black}\href{https://doi.org/10.1103/PhysRevX.11.021053}{\detokenize{10.1103/PhysRevX.11.021053}}}.

\bibitem[Abbott \em{et~al.}(2021{\natexlab{b}})Abbott
  et~al.]{LIGOScientific:2021usb}
Abbott, R.; others.
\newblock {GWTC-2.1: Deep Extended Catalog of Compact Binary Coalescences
  Observed by LIGO and Virgo During the First Half of the Third Observing Run}
  {\bf 2021}.
\newblock  \href{http://xxx.lanl.gov/abs/2108.01045}{{\normalfont
  [arXiv:gr-qc/2108.01045]}}.

\bibitem[Abbott \em{et~al.}(2021{\natexlab{c}})Abbott et~al.]{KAGRA:2021kbb}
Abbott, R.; others.
\newblock {Upper limits on the isotropic gravitational-wave background from
  Advanced LIGO and Advanced Virgo\textquoteright{}s third observing run}.
\newblock {\em Phys. Rev. D} {\bf 2021}, {\em 104},~022004,
  \href{http://xxx.lanl.gov/abs/2101.12130}{{\normalfont
  [arXiv:gr-qc/2101.12130]}}.
\newblock
  doi:{\changeurlcolor{black}\href{https://doi.org/10.1103/PhysRevD.104.022004}{\detokenize{10.1103/PhysRevD.104.022004}}}.

\bibitem[Abbott \em{et~al.}(2021{\natexlab{d}})Abbott et~al.]{KAGRA:2021mth}
Abbott, R.; others.
\newblock {Search for anisotropic gravitational-wave backgrounds using data
  from Advanced LIGO and Advanced Virgo\textquoteright{}s first three observing
  runs}.
\newblock {\em Phys. Rev. D} {\bf 2021}, {\em 104},~022005,
  \href{http://xxx.lanl.gov/abs/2103.08520}{{\normalfont
  [arXiv:gr-qc/2103.08520]}}.
\newblock
  doi:{\changeurlcolor{black}\href{https://doi.org/10.1103/PhysRevD.104.022005}{\detokenize{10.1103/PhysRevD.104.022005}}}.

\bibitem[Christensen(2019)]{Christensen:2018iqi}
Christensen, N.
\newblock {Stochastic Gravitational Wave Backgrounds}.
\newblock {\em Rept. Prog. Phys.} {\bf 2019}, {\em 82},~016903,
  \href{http://xxx.lanl.gov/abs/1811.08797}{{\normalfont
  [arXiv:gr-qc/1811.08797]}}.
\newblock
  doi:{\changeurlcolor{black}\href{https://doi.org/10.1088/1361-6633/aae6b5}{\detokenize{10.1088/1361-6633/aae6b5}}}.

\bibitem[Regimbau(2011)]{Regimbau:2011rp}
Regimbau, T.
\newblock {The astrophysical gravitational wave stochastic background}.
\newblock {\em Res. Astron. Astrophys.} {\bf 2011}, {\em 11},~369--390,
  \href{http://xxx.lanl.gov/abs/1101.2762}{{\normalfont
  [arXiv:astro-ph.CO/1101.2762]}}.
\newblock
  doi:{\changeurlcolor{black}\href{https://doi.org/10.1088/1674-4527/11/4/001}{\detokenize{10.1088/1674-4527/11/4/001}}}.

\bibitem[Rosado(2011)]{Rosado:2011kv}
Rosado, P.A.
\newblock {Gravitational wave background from binary systems}.
\newblock {\em Phys. Rev. D} {\bf 2011}, {\em 84},~084004,
  \href{http://xxx.lanl.gov/abs/1106.5795}{{\normalfont
  [arXiv:gr-qc/1106.5795]}}.
\newblock
  doi:{\changeurlcolor{black}\href{https://doi.org/10.1103/PhysRevD.84.084004}{\detokenize{10.1103/PhysRevD.84.084004}}}.

\bibitem[Marassi \em{et~al.}(2011)Marassi, Schneider, Corvino, Ferrari, and
  Portegies~Zwart]{Marassi:2011si}
Marassi, S.; Schneider, R.; Corvino, G.; Ferrari, V.; Portegies~Zwart, S.
\newblock {Imprint of the merger and ring-down on the gravitational wave
  background from black hole binaries coalescence}.
\newblock {\em Phys. Rev. D} {\bf 2011}, {\em 84},~124037,
  \href{http://xxx.lanl.gov/abs/1111.6125}{{\normalfont
  [arXiv:astro-ph.CO/1111.6125]}}.
\newblock
  doi:{\changeurlcolor{black}\href{https://doi.org/10.1103/PhysRevD.84.124037}{\detokenize{10.1103/PhysRevD.84.124037}}}.

\bibitem[Zhu \em{et~al.}(2011)Zhu, Howell, Regimbau, Blair, and
  Zhu]{Zhu:2011bd}
Zhu, X.J.; Howell, E.; Regimbau, T.; Blair, D.; Zhu, Z.H.
\newblock {Stochastic Gravitational Wave Background from Coalescing Binary
  Black Holes}.
\newblock {\em Astrophys. J.} {\bf 2011}, {\em 739},~86,
  \href{http://xxx.lanl.gov/abs/1104.3565}{{\normalfont
  [arXiv:gr-qc/1104.3565]}}.
\newblock
  doi:{\changeurlcolor{black}\href{https://doi.org/10.1088/0004-637X/739/2/86}{\detokenize{10.1088/0004-637X/739/2/86}}}.

\bibitem[Zhu \em{et~al.}(2013)Zhu, Howell, Blair, and Zhu]{Zhu:2012xw}
Zhu, X.J.; Howell, E.J.; Blair, D.G.; Zhu, Z.H.
\newblock {On the gravitational wave background from compact binary
  coalescences in the band of ground-based interferometers}.
\newblock {\em Mon. Not. Roy. Astron. Soc.} {\bf 2013}, {\em 431},~882--899,
  \href{http://xxx.lanl.gov/abs/1209.0595}{{\normalfont
  [arXiv:gr-qc/1209.0595]}}.
\newblock
  doi:{\changeurlcolor{black}\href{https://doi.org/10.1093/mnras/stt207}{\detokenize{10.1093/mnras/stt207}}}.

\bibitem[Wu \em{et~al.}(2012)Wu, Mandic, and Regimbau]{Wu:2011ac}
Wu, C.; Mandic, V.; Regimbau, T.
\newblock {Accessibility of the Gravitational-Wave Background due to Binary
  Coalescences to Second and Third Generation Gravitational-Wave Detectors}.
\newblock {\em Phys. Rev. D} {\bf 2012}, {\em 85},~104024,
  \href{http://xxx.lanl.gov/abs/1112.1898}{{\normalfont
  [arXiv:gr-qc/1112.1898]}}.
\newblock
  doi:{\changeurlcolor{black}\href{https://doi.org/10.1103/PhysRevD.85.104024}{\detokenize{10.1103/PhysRevD.85.104024}}}.

\bibitem[{Kowalska-Leszczynska, I.} \em{et~al.}(2015){Kowalska-Leszczynska,
  I.}, {Regimbau, T.}, {Bulik, T.}, {Dominik, M.}, and {Belczynski,
  K.}]{KowalskaLeszczynska2015}
{Kowalska-Leszczynska, I.}.; {Regimbau, T.}.; {Bulik, T.}.; {Dominik, M.}.;
  {Belczynski, K.}.
\newblock Effect of metallicity on the gravitational-wave signal from the
  cosmological population of compact binary coalescences.
\newblock {\em A\&A} {\bf 2015}, {\em 574},~A58.
\newblock
  doi:{\changeurlcolor{black}\href{https://doi.org/10.1051/0004-6361/201424417}{\detokenize{10.1051/0004-6361/201424417}}}.

\bibitem[Abbott \em{et~al.}(2016)Abbott et~al.]{TheLIGOScientific:2016wyq}
Abbott, B.; others.
\newblock {GW150914: Implications for the stochastic gravitational wave
  background from binary black holes}.
\newblock {\em Phys. Rev. Lett.} {\bf 2016}, {\em 116},~131102,
  \href{http://xxx.lanl.gov/abs/1602.03847}{{\normalfont
  [arXiv:gr-qc/1602.03847]}}.
\newblock
  doi:{\changeurlcolor{black}\href{https://doi.org/10.1103/PhysRevLett.116.131102}{\detokenize{10.1103/PhysRevLett.116.131102}}}.

\bibitem[Abbott \em{et~al.}(2018)Abbott et~al.]{Abbott:2017xzg}
Abbott, B.P.; others.
\newblock {GW170817: Implications for the Stochastic Gravitational-Wave
  Background from Compact Binary Coalescences}.
\newblock {\em Phys. Rev. Lett.} {\bf 2018}, {\em 120},~091101,
  \href{http://xxx.lanl.gov/abs/1710.05837}{{\normalfont
  [arXiv:gr-qc/1710.05837]}}.
\newblock
  doi:{\changeurlcolor{black}\href{https://doi.org/10.1103/PhysRevLett.120.091101}{\detokenize{10.1103/PhysRevLett.120.091101}}}.

\bibitem[P\'erigois \em{et~al.}(2021)P\'erigois, Belczynski, Bulik, and
  Regimbau]{Perigois:2020ymr}
P\'erigois, C.; Belczynski, C.; Bulik, T.; Regimbau, T.
\newblock {StarTrack predictions of the stochastic gravitational-wave
  background from compact binary mergers}.
\newblock {\em Phys. Rev. D} {\bf 2021}, {\em 103},~043002,
  \href{http://xxx.lanl.gov/abs/2008.04890}{{\normalfont
  [arXiv:astro-ph.CO/2008.04890]}}.
\newblock
  doi:{\changeurlcolor{black}\href{https://doi.org/10.1103/PhysRevD.103.043002}{\detokenize{10.1103/PhysRevD.103.043002}}}.

\bibitem[Contaldi(2017)]{Contaldi:2016koz}
Contaldi, C.R.
\newblock {Anisotropies of Gravitational Wave Backgrounds: A Line Of Sight
  Approach}.
\newblock {\em Phys. Lett. B} {\bf 2017}, {\em 771},~9--12,
  \href{http://xxx.lanl.gov/abs/1609.08168}{{\normalfont
  [arXiv:astro-ph.CO/1609.08168]}}.
\newblock
  doi:{\changeurlcolor{black}\href{https://doi.org/10.1016/j.physletb.2017.05.020}{\detokenize{10.1016/j.physletb.2017.05.020}}}.

\bibitem[Cusin \em{et~al.}(2017)Cusin, Pitrou, and Uzan]{Cusin:2017fwz}
Cusin, G.; Pitrou, C.; Uzan, J.P.
\newblock {Anisotropy of the astrophysical gravitational wave background:
  Analytic expression of the angular power spectrum and correlation with
  cosmological observations}.
\newblock {\em Phys. Rev. D} {\bf 2017}, {\em 96},~103019,
  \href{http://xxx.lanl.gov/abs/1704.06184}{{\normalfont
  [arXiv:astro-ph.CO/1704.06184]}}.
\newblock
  doi:{\changeurlcolor{black}\href{https://doi.org/10.1103/PhysRevD.96.103019}{\detokenize{10.1103/PhysRevD.96.103019}}}.

\bibitem[Cusin \em{et~al.}(2018)Cusin, Pitrou, and Uzan]{Cusin:2017mjm}
Cusin, G.; Pitrou, C.; Uzan, J.P.
\newblock {The signal of the gravitational wave background and the angular
  correlation of its energy density}.
\newblock {\em Phys. Rev. D} {\bf 2018}, {\em 97},~123527,
  \href{http://xxx.lanl.gov/abs/1711.11345}{{\normalfont
  [arXiv:astro-ph.CO/1711.11345]}}.
\newblock
  doi:{\changeurlcolor{black}\href{https://doi.org/10.1103/PhysRevD.97.123527}{\detokenize{10.1103/PhysRevD.97.123527}}}.

\bibitem[Jenkins \em{et~al.}(2018)Jenkins, Sakellariadou, Regimbau, and
  Slezak]{Jenkins:2018uac}
Jenkins, A.C.; Sakellariadou, M.; Regimbau, T.; Slezak, E.
\newblock {Anisotropies in the astrophysical gravitational-wave background:
  Predictions for the detection of compact binaries by LIGO and Virgo}.
\newblock {\em Phys. Rev. D} {\bf 2018}, {\em 98},~063501,
  \href{http://xxx.lanl.gov/abs/1806.01718}{{\normalfont
  [arXiv:astro-ph.CO/1806.01718]}}.
\newblock
  doi:{\changeurlcolor{black}\href{https://doi.org/10.1103/PhysRevD.98.063501}{\detokenize{10.1103/PhysRevD.98.063501}}}.

\bibitem[Jenkins \em{et~al.}(2019)Jenkins, O'Shaughnessy, Sakellariadou, and
  Wysocki]{Jenkins:2018kxc}
Jenkins, A.C.; O'Shaughnessy, R.; Sakellariadou, M.; Wysocki, D.
\newblock {Anisotropies in the astrophysical gravitational-wave background: The
  impact of black hole distributions}.
\newblock {\em Phys. Rev. Lett.} {\bf 2019}, {\em 122},~111101,
  \href{http://xxx.lanl.gov/abs/1810.13435}{{\normalfont
  [arXiv:astro-ph.CO/1810.13435]}}.
\newblock
  doi:{\changeurlcolor{black}\href{https://doi.org/10.1103/PhysRevLett.122.111101}{\detokenize{10.1103/PhysRevLett.122.111101}}}.

\bibitem[Cusin \em{et~al.}(2018)Cusin, Dvorkin, Pitrou, and
  Uzan]{Cusin:2018rsq}
Cusin, G.; Dvorkin, I.; Pitrou, C.; Uzan, J.P.
\newblock {First predictions of the angular power spectrum of the astrophysical
  gravitational wave background}.
\newblock {\em Phys. Rev. Lett.} {\bf 2018}, {\em 120},~231101,
  \href{http://xxx.lanl.gov/abs/1803.03236}{{\normalfont
  [arXiv:astro-ph.CO/1803.03236]}}.
\newblock
  doi:{\changeurlcolor{black}\href{https://doi.org/10.1103/PhysRevLett.120.231101}{\detokenize{10.1103/PhysRevLett.120.231101}}}.

\bibitem[Cusin \em{et~al.}(2019)Cusin, Dvorkin, Pitrou, and
  Uzan]{Cusin:2019jpv}
Cusin, G.; Dvorkin, I.; Pitrou, C.; Uzan, J.P.
\newblock {Properties of the stochastic astrophysical gravitational wave
  background: astrophysical sources dependencies}.
\newblock {\em Phys. Rev. D} {\bf 2019}, {\em 100},~063004,
  \href{http://xxx.lanl.gov/abs/1904.07797}{{\normalfont
  [arXiv:astro-ph.CO/1904.07797]}}.
\newblock
  doi:{\changeurlcolor{black}\href{https://doi.org/10.1103/PhysRevD.100.063004}{\detokenize{10.1103/PhysRevD.100.063004}}}.

\bibitem[Bertacca \em{et~al.}(2020)Bertacca, Ricciardone, Bellomo, Jenkins,
  Matarrese, Raccanelli, Regimbau, and Sakellariadou]{Bertacca:2019fnt}
Bertacca, D.; Ricciardone, A.; Bellomo, N.; Jenkins, A.C.; Matarrese, S.;
  Raccanelli, A.; Regimbau, T.; Sakellariadou, M.
\newblock {Projection effects on the observed angular spectrum of the
  astrophysical stochastic gravitational wave background}.
\newblock {\em Phys. Rev. D} {\bf 2020}, {\em 101},~103513,
  \href{http://xxx.lanl.gov/abs/1909.11627}{{\normalfont
  [arXiv:astro-ph.CO/1909.11627]}}.
\newblock
  doi:{\changeurlcolor{black}\href{https://doi.org/10.1103/PhysRevD.101.103513}{\detokenize{10.1103/PhysRevD.101.103513}}}.

\bibitem[Pitrou \em{et~al.}(2020)Pitrou, Cusin, and Uzan]{Pitrou:2019rjz}
Pitrou, C.; Cusin, G.; Uzan, J.P.
\newblock {Unified view of anisotropies in the astrophysical gravitational-wave
  background}.
\newblock {\em Phys. Rev. D} {\bf 2020}, {\em 101},~081301,
  \href{http://xxx.lanl.gov/abs/1910.04645}{{\normalfont
  [arXiv:astro-ph.CO/1910.04645]}}.
\newblock
  doi:{\changeurlcolor{black}\href{https://doi.org/10.1103/PhysRevD.101.081301}{\detokenize{10.1103/PhysRevD.101.081301}}}.

\bibitem[Capurri \em{et~al.}(2021)Capurri, Lapi, Baccigalupi, Boco, Scelfo, and
  Ronconi]{Capurri:2021zli}
Capurri, G.; Lapi, A.; Baccigalupi, C.; Boco, L.; Scelfo, G.; Ronconi, T.
\newblock {Intensity and anisotropies of the stochastic gravitational wave
  background from merging compact binaries in galaxies}.
\newblock {\em JCAP} {\bf 2021}, {\em 11},~032,
  \href{http://xxx.lanl.gov/abs/2103.12037}{{\normalfont
  [arXiv:gr-qc/2103.12037]}}.
\newblock
  doi:{\changeurlcolor{black}\href{https://doi.org/10.1088/1475-7516/2021/11/032}{\detokenize{10.1088/1475-7516/2021/11/032}}}.

\bibitem[Bellomo \em{et~al.}(2021)Bellomo, Bertacca, Jenkins, Matarrese,
  Raccanelli, Regimbau, Ricciardone, and Sakellariadou]{Bellomo:2021mer}
Bellomo, N.; Bertacca, D.; Jenkins, A.C.; Matarrese, S.; Raccanelli, A.;
  Regimbau, T.; Ricciardone, A.; Sakellariadou, M.
\newblock {CLASS\_GWB: robust modeling of the astrophysical gravitational wave
  background anisotropies} {\bf 2021}.
\newblock  \href{http://xxx.lanl.gov/abs/2110.15059}{{\normalfont
  [arXiv:gr-qc/2110.15059]}}.

\bibitem[Cusin \em{et~al.}(2020)Cusin, Dvorkin, Pitrou, and
  Uzan]{Cusin:2019jhg}
Cusin, G.; Dvorkin, I.; Pitrou, C.; Uzan, J.P.
\newblock {Stochastic gravitational wave background anisotropies in the mHz
  band: astrophysical dependencies}.
\newblock {\em Mon. Not. Roy. Astron. Soc.} {\bf 2020}, {\em 493},~L1--L5,
  \href{http://xxx.lanl.gov/abs/1904.07757}{{\normalfont
  [arXiv:astro-ph.CO/1904.07757]}}.
\newblock
  doi:{\changeurlcolor{black}\href{https://doi.org/10.1093/mnrasl/slz182}{\detokenize{10.1093/mnrasl/slz182}}}.

\bibitem[Abbott \em{et~al.}(2019)Abbott et~al.]{LIGOScientific:2019gaw}
Abbott, B.P.; others.
\newblock {Directional limits on persistent gravitational waves using data from
  Advanced LIGO's first two observing runs}.
\newblock {\em Phys. Rev. D} {\bf 2019}, {\em 100},~062001,
  \href{http://xxx.lanl.gov/abs/1903.08844}{{\normalfont
  [arXiv:gr-qc/1903.08844]}}.
\newblock
  doi:{\changeurlcolor{black}\href{https://doi.org/10.1103/PhysRevD.100.062001}{\detokenize{10.1103/PhysRevD.100.062001}}}.

\bibitem[Abbott \em{et~al.}(2017)Abbott et~al.]{LIGOScientific:2016nwa}
Abbott, B.P.; others.
\newblock {Directional Limits on Persistent Gravitational Waves from Advanced
  LIGO\textquoteright{}s First Observing Run}.
\newblock {\em Phys. Rev. Lett.} {\bf 2017}, {\em 118},~121102,
  \href{http://xxx.lanl.gov/abs/1612.02030}{{\normalfont
  [arXiv:gr-qc/1612.02030]}}.
\newblock
  doi:{\changeurlcolor{black}\href{https://doi.org/10.1103/PhysRevLett.118.121102}{\detokenize{10.1103/PhysRevLett.118.121102}}}.

\bibitem[Renzini and Contaldi(2019)]{Renzini:2018nee}
Renzini, A.I.; Contaldi, C.R.
\newblock {Gravitational Wave Background Sky Maps from Advanced LIGO O1 Data}.
\newblock {\em Phys. Rev. Lett.} {\bf 2019}, {\em 122},~081102,
  \href{http://xxx.lanl.gov/abs/1811.12922}{{\normalfont
  [arXiv:astro-ph.CO/1811.12922]}}.
\newblock
  doi:{\changeurlcolor{black}\href{https://doi.org/10.1103/PhysRevLett.122.081102}{\detokenize{10.1103/PhysRevLett.122.081102}}}.

\bibitem[Mentasti and Peloso(2021)]{Mentasti:2020yyd}
Mentasti, G.; Peloso, M.
\newblock {ET sensitivity to the anisotropic Stochastic Gravitational Wave
  Background}.
\newblock {\em JCAP} {\bf 2021}, {\em 03},~080,
  \href{http://xxx.lanl.gov/abs/2010.00486}{{\normalfont
  [arXiv:astro-ph.CO/2010.00486]}}.
\newblock
  doi:{\changeurlcolor{black}\href{https://doi.org/10.1088/1475-7516/2021/03/080}{\detokenize{10.1088/1475-7516/2021/03/080}}}.

\bibitem[Thrane \em{et~al.}(2009)Thrane, Ballmer, Romano, Mitra, Talukder,
  Bose, and Mandic]{Thrane:2009}
Thrane, E.; Ballmer, S.; Romano, J.D.; Mitra, S.; Talukder, D.; Bose, S.;
  Mandic, V.
\newblock Probing the anisotropies of a stochastic gravitational-wave
  background using a network of ground-based laser interferometers.
\newblock {\em Physical Review D} {\bf 2009}, {\em 80}.
\newblock
  doi:{\changeurlcolor{black}\href{https://doi.org/10.1103/physrevd.80.122002}{\detokenize{10.1103/physrevd.80.122002}}}.

\bibitem[Gair \em{et~al.}(2014)Gair, Romano, Taylor, and
  Mingarelli]{Gair:2014rwa}
Gair, J.; Romano, J.D.; Taylor, S.; Mingarelli, C.M.F.
\newblock {Mapping gravitational-wave backgrounds using methods from CMB
  analysis: Application to pulsar timing arrays}.
\newblock {\em Phys. Rev. D} {\bf 2014}, {\em 90},~082001,
  \href{http://xxx.lanl.gov/abs/1406.4664}{{\normalfont
  [arXiv:gr-qc/1406.4664]}}.
\newblock
  doi:{\changeurlcolor{black}\href{https://doi.org/10.1103/PhysRevD.90.082001}{\detokenize{10.1103/PhysRevD.90.082001}}}.

\bibitem[Romano \em{et~al.}(2015)Romano, Taylor, Cornish, Gair, Mingarelli, and
  van Haasteren]{Romano:2015uma}
Romano, J.D.; Taylor, S.R.; Cornish, N.J.; Gair, J.; Mingarelli, C.M.F.; van
  Haasteren, R.
\newblock {Phase-coherent mapping of gravitational-wave backgrounds using
  ground-based laser interferometers}.
\newblock {\em Phys. Rev. D} {\bf 2015}, {\em 92},~042003,
  \href{http://xxx.lanl.gov/abs/1505.07179}{{\normalfont
  [arXiv:gr-qc/1505.07179]}}.
\newblock
  doi:{\changeurlcolor{black}\href{https://doi.org/10.1103/PhysRevD.92.042003}{\detokenize{10.1103/PhysRevD.92.042003}}}.

\bibitem[Ain \em{et~al.}(2018)Ain, Suresh, and Mitra]{Ain:2018zvo}
Ain, A.; Suresh, J.; Mitra, S.
\newblock {Very fast stochastic gravitational wave background map making using
  folded data}.
\newblock {\em Phys. Rev. D} {\bf 2018}, {\em 98},~024001,
  \href{http://xxx.lanl.gov/abs/1803.08285}{{\normalfont
  [arXiv:gr-qc/1803.08285]}}.
\newblock
  doi:{\changeurlcolor{black}\href{https://doi.org/10.1103/PhysRevD.98.024001}{\detokenize{10.1103/PhysRevD.98.024001}}}.

\bibitem[Renzini and Contaldi(2018)]{Renzini:2018vkx}
Renzini, A.I.; Contaldi, C.R.
\newblock {Mapping Incoherent Gravitational Wave Backgrounds}.
\newblock {\em Mon. Not. Roy. Astron. Soc.} {\bf 2018}, {\em 481},~4650--4661,
  \href{http://xxx.lanl.gov/abs/1806.11360}{{\normalfont
  [arXiv:astro-ph.IM/1806.11360]}}.
\newblock
  doi:{\changeurlcolor{black}\href{https://doi.org/10.1093/mnras/sty2546}{\detokenize{10.1093/mnras/sty2546}}}.

\bibitem[Conneely \em{et~al.}(2019)Conneely, Jaffe, and
  Mingarelli]{Conneely:2018wis}
Conneely, C.; Jaffe, A.H.; Mingarelli, C.M.F.
\newblock {On the Amplitude and Stokes Parameters of a Stochastic
  Gravitational-Wave Background}.
\newblock {\em Mon. Not. Roy. Astron. Soc.} {\bf 2019}, {\em 487},~562--579,
  \href{http://xxx.lanl.gov/abs/1808.05920}{{\normalfont
  [arXiv:astro-ph.CO/1808.05920]}}.
\newblock
  doi:{\changeurlcolor{black}\href{https://doi.org/10.1093/mnras/stz1022}{\detokenize{10.1093/mnras/stz1022}}}.

\bibitem[Hotinli \em{et~al.}(2019)Hotinli, Kamionkowski, and
  Jaffe]{Hotinli:2019tpc}
Hotinli, S.C.; Kamionkowski, M.; Jaffe, A.H.
\newblock {The search for anisotropy in the gravitational-wave background with
  pulsar-timing arrays}.
\newblock {\em Open J. Astrophys.} {\bf 2019}, {\em 2},~8,
  \href{http://xxx.lanl.gov/abs/1904.05348}{{\normalfont
  [arXiv:astro-ph.CO/1904.05348]}}.
\newblock
  doi:{\changeurlcolor{black}\href{https://doi.org/10.21105/astro.1904.05348}{\detokenize{10.21105/astro.1904.05348}}}.

\bibitem[Alonso \em{et~al.}(2020{\natexlab{a}})Alonso, Contaldi, Cusin,
  Ferreira, and Renzini]{Alonso:2020rar}
Alonso, D.; Contaldi, C.R.; Cusin, G.; Ferreira, P.G.; Renzini, A.I.
\newblock {Noise angular power spectrum of gravitational wave background
  experiments}.
\newblock {\em Phys. Rev. D} {\bf 2020}, {\em 101},~124048,
  \href{http://xxx.lanl.gov/abs/2005.03001}{{\normalfont
  [arXiv:astro-ph.CO/2005.03001]}}.
\newblock
  doi:{\changeurlcolor{black}\href{https://doi.org/10.1103/PhysRevD.101.124048}{\detokenize{10.1103/PhysRevD.101.124048}}}.

\bibitem[Alonso \em{et~al.}(2020{\natexlab{b}})Alonso, Cusin, Ferreira, and
  Pitrou]{Alonso:2020mva}
Alonso, D.; Cusin, G.; Ferreira, P.G.; Pitrou, C.
\newblock {Detecting the anisotropic astrophysical gravitational wave
  background in the presence of shot noise through cross-correlations}.
\newblock {\em Phys. Rev. D} {\bf 2020}, {\em 102},~023002,
  \href{http://xxx.lanl.gov/abs/2002.02888}{{\normalfont
  [arXiv:astro-ph.CO/2002.02888]}}.
\newblock
  doi:{\changeurlcolor{black}\href{https://doi.org/10.1103/PhysRevD.102.023002}{\detokenize{10.1103/PhysRevD.102.023002}}}.

\bibitem[Jenkins \em{et~al.}(2019)Jenkins, Romano, and
  Sakellariadou]{Jenkins:2019nks}
Jenkins, A.C.; Romano, J.D.; Sakellariadou, M.
\newblock {Estimating the angular power spectrum of the gravitational-wave
  background in the presence of shot noise}.
\newblock {\em Phys. Rev. D} {\bf 2019}, {\em 100},~083501,
  \href{http://xxx.lanl.gov/abs/1907.06642}{{\normalfont
  [arXiv:astro-ph.CO/1907.06642]}}.
\newblock
  doi:{\changeurlcolor{black}\href{https://doi.org/10.1103/PhysRevD.100.083501}{\detokenize{10.1103/PhysRevD.100.083501}}}.

\bibitem[Jenkins and Sakellariadou(2019)]{Jenkins:2019uzp}
Jenkins, A.C.; Sakellariadou, M.
\newblock {Shot noise in the astrophysical gravitational-wave background}.
\newblock {\em Phys. Rev. D} {\bf 2019}, {\em 100},~063508,
  \href{http://xxx.lanl.gov/abs/1902.07719}{{\normalfont
  [arXiv:astro-ph.CO/1902.07719]}}.
\newblock
  doi:{\changeurlcolor{black}\href{https://doi.org/10.1103/PhysRevD.100.063508}{\detokenize{10.1103/PhysRevD.100.063508}}}.

\bibitem[Ca\~nas Herrera \em{et~al.}(2020)Ca\~nas Herrera, Contigiani, and
  Vardanyan]{Canas-Herrera:2019npr}
Ca\~nas Herrera, G.; Contigiani, O.; Vardanyan, V.
\newblock {Cross-correlation of the astrophysical gravitational-wave background
  with galaxy clustering}.
\newblock {\em Phys. Rev. D} {\bf 2020}, {\em 102},~043513,
  \href{http://xxx.lanl.gov/abs/1910.08353}{{\normalfont
  [arXiv:astro-ph.CO/1910.08353]}}.
\newblock
  doi:{\changeurlcolor{black}\href{https://doi.org/10.1103/PhysRevD.102.043513}{\detokenize{10.1103/PhysRevD.102.043513}}}.

\bibitem[Yang \em{et~al.}(2020)Yang, Mandic, Scarlata, and
  Banagiri]{Yang:2020usq}
Yang, K.Z.; Mandic, V.; Scarlata, C.; Banagiri, S.
\newblock {Searching for Cross-Correlation Between Stochastic Gravitational
  Wave Background and Galaxy Number Counts}.
\newblock {\em Mon. Not. Roy. Astron. Soc.} {\bf 2020}, {\em 500},~1666--1672,
  \href{http://xxx.lanl.gov/abs/2007.10456}{{\normalfont
  [arXiv:astro-ph.CO/2007.10456]}}.
\newblock
  doi:{\changeurlcolor{black}\href{https://doi.org/10.1093/mnras/staa3159}{\detokenize{10.1093/mnras/staa3159}}}.

\bibitem[Mukherjee and Silk(2020)]{Mukherjee:2019oma}
Mukherjee, S.; Silk, J.
\newblock {Time-dependence of the astrophysical stochastic gravitational wave
  background}.
\newblock {\em Mon. Not. Roy. Astron. Soc.} {\bf 2020}, {\em 491},~4690--4701,
  \href{http://xxx.lanl.gov/abs/1912.07657}{{\normalfont
  [arXiv:gr-qc/1912.07657]}}.
\newblock
  doi:{\changeurlcolor{black}\href{https://doi.org/10.1093/mnras/stz3226}{\detokenize{10.1093/mnras/stz3226}}}.

\bibitem[Ricciardone \em{et~al.}(2021)Ricciardone, Dall'Armi, Bartolo,
  Bertacca, Liguori, and Matarrese]{Ricciardone:2021kel}
Ricciardone, A.; Dall'Armi, L.V.; Bartolo, N.; Bertacca, D.; Liguori, M.;
  Matarrese, S.
\newblock {Cross-Correlating Astrophysical and Cosmological Gravitational Wave
  Backgrounds with the Cosmic Microwave Background}.
\newblock {\em Phys. Rev. Lett.} {\bf 2021}, {\em 127},~271301,
  \href{http://xxx.lanl.gov/abs/2106.02591}{{\normalfont
  [arXiv:astro-ph.CO/2106.02591]}}.
\newblock
  doi:{\changeurlcolor{black}\href{https://doi.org/10.1103/PhysRevLett.127.271301}{\detokenize{10.1103/PhysRevLett.127.271301}}}.

\bibitem[Braglia and Kuroyanagi(2021)]{Braglia:2021fxn}
Braglia, M.; Kuroyanagi, S.
\newblock {Probing prerecombination physics by the cross-correlation of
  stochastic gravitational waves and CMB anisotropies}.
\newblock {\em Phys. Rev. D} {\bf 2021}, {\em 104},~123547,
  \href{http://xxx.lanl.gov/abs/2106.03786}{{\normalfont
  [arXiv:astro-ph.CO/2106.03786]}}.
\newblock
  doi:{\changeurlcolor{black}\href{https://doi.org/10.1103/PhysRevD.104.123547}{\detokenize{10.1103/PhysRevD.104.123547}}}.

\bibitem[Ade \em{et~al.}(2021)Ade et~al.]{BICEP:2021xfz}
Ade, P.A.R.; others.
\newblock {Improved Constraints on Primordial Gravitational Waves using Planck,
  WMAP, and BICEP/Keck Observations through the 2018 Observing Season}.
\newblock {\em Phys. Rev. Lett.} {\bf 2021}, {\em 127},~151301,
  \href{http://xxx.lanl.gov/abs/2110.00483}{{\normalfont
  [arXiv:astro-ph.CO/2110.00483]}}.
\newblock
  doi:{\changeurlcolor{black}\href{https://doi.org/10.1103/PhysRevLett.127.151301}{\detokenize{10.1103/PhysRevLett.127.151301}}}.

\bibitem[Han \em{et~al.}(2021)Han et~al.]{ACT:2020goa}
Han, D.; others.
\newblock {The Atacama Cosmology Telescope: delensed power spectra and
  parameters}.
\newblock {\em JCAP} {\bf 2021}, {\em 01},~031,
  \href{http://xxx.lanl.gov/abs/2007.14405}{{\normalfont
  [arXiv:astro-ph.CO/2007.14405]}}.
\newblock
  doi:{\changeurlcolor{black}\href{https://doi.org/10.1088/1475-7516/2021/01/031}{\detokenize{10.1088/1475-7516/2021/01/031}}}.

\bibitem[Adachi \em{et~al.}(2020)Adachi et~al.]{POLARBEAR:2019kzz}
Adachi, S.; others.
\newblock {A Measurement of the Degree Scale CMB $B$-mode Angular Power
  Spectrum with POLARBEAR}.
\newblock {\em Astrophys. J.} {\bf 2020}, {\em 897},~55,
  \href{http://xxx.lanl.gov/abs/1910.02608}{{\normalfont
  [arXiv:astro-ph.CO/1910.02608]}}.
\newblock
  doi:{\changeurlcolor{black}\href{https://doi.org/10.3847/1538-4357/ab8f24}{\detokenize{10.3847/1538-4357/ab8f24}}}.

\bibitem[Guidi \em{et~al.}(2021)Guidi, Rubiño-Martín, Pelaez-Santos,
  Génova-Santos, Ashdown, Barreiro, Bilbao-Ahedo, Harper, and
  Watson]{quijote:2021}
Guidi, F.; Rubiño-Martín, J.A.; Pelaez-Santos, A.E.; Génova-Santos, R.T.;
  Ashdown, M.; Barreiro, R.B.; Bilbao-Ahedo, J.D.; Harper, S.E.; Watson, R.A.
\newblock The PICASSO map-making code: application to a simulation of the
  QUIJOTE northern sky survey.
\newblock {\em Monthly Notices of the Royal Astronomical Society} {\bf 2021},
  {\em 507},~3707–3725.
\newblock
  doi:{\changeurlcolor{black}\href{https://doi.org/10.1093/mnras/stab2422}{\detokenize{10.1093/mnras/stab2422}}}.

\bibitem[Dutcher \em{et~al.}(2021)Dutcher et~al.]{SPT-3G:2021eoc}
Dutcher, D.; others.
\newblock {Measurements of the E-mode polarization and temperature-E-mode
  correlation of the CMB from SPT-3G 2018 data}.
\newblock {\em Phys. Rev. D} {\bf 2021}, {\em 104},~022003,
  \href{http://xxx.lanl.gov/abs/2101.01684}{{\normalfont
  [arXiv:astro-ph.CO/2101.01684]}}.
\newblock
  doi:{\changeurlcolor{black}\href{https://doi.org/10.1103/PhysRevD.104.022003}{\detokenize{10.1103/PhysRevD.104.022003}}}.

\bibitem[Dahal \em{et~al.}(2021)Dahal et~al.]{Dahal:2021uig}
Dahal, S.; others.
\newblock {Four-year Cosmology Large Angular Scale Surveyor (CLASS)
  Observations: On-sky Receiver Performance at 40, 90, 150, and 220 GHz
  Frequency Bands} {\bf 2021}.
\newblock  \href{http://xxx.lanl.gov/abs/2107.08022}{{\normalfont
  [arXiv:astro-ph.IM/2107.08022]}}.

\bibitem[Addamo \em{et~al.}(2021)Addamo et~al.]{LSPE:2020uos}
Addamo, G.; others.
\newblock {The large scale polarization explorer (LSPE) for CMB measurements:
  performance forecast}.
\newblock {\em JCAP} {\bf 2021}, {\em 08},~008,
  \href{http://xxx.lanl.gov/abs/2008.11049}{{\normalfont
  [arXiv:astro-ph.IM/2008.11049]}}.
\newblock
  doi:{\changeurlcolor{black}\href{https://doi.org/10.1088/1475-7516/2021/08/008}{\detokenize{10.1088/1475-7516/2021/08/008}}}.

\bibitem[Ade \em{et~al.}(2019)Ade et~al.]{Ade2019}
Ade, P.; others.
\newblock The Simons Observatory: Science goals and forecasts.
\newblock {\em Journal of Cosmology and Astroparticle Physics} {\bf 2019}, {\em
  2019}.
\newblock
  doi:{\changeurlcolor{black}\href{https://doi.org/10.1088/1475-7516/2019/02/056}{\detokenize{10.1088/1475-7516/2019/02/056}}}.

\bibitem[Abazajian \em{et~al.}(2020)Abazajian et~al.]{CMB-S4:2020lpa}
Abazajian, K.; others.
\newblock {CMB-S4: Forecasting Constraints on Primordial Gravitational Waves}
  {\bf 2020}.
\newblock  \href{http://xxx.lanl.gov/abs/2008.12619}{{\normalfont
  [arXiv:astro-ph.CO/2008.12619]}}.

\bibitem[Ganga \em{et~al.}(2019)Ganga, Baccigalupi, Bouchet, Brown, Challinor,
  Komatsu, Martínez-Gonzalez, Mennella, Mohr, Rubiño-Martín, Bersanelli,
  Barreiro, Lopez-Callado, Colombo, Delabrouille, Errard, Hamilton, Jones,
  Krachmalnicoff, Montier, Natoli, Piacentini, Poletti, Stompor, Taylor,
  Vielva, and Vittorio]{Ganga2019European}
Ganga, K.; Baccigalupi, C.; Bouchet, F.; Brown, M.; Challinor, A.; Komatsu, E.;
  Martínez-Gonzalez, E.; Mennella, D.; Mohr, J.; Rubiño-Martín, J.A.;
  Bersanelli, M.; Barreiro, B.; Lopez-Callado, E.d.l.H.; Colombo, L.;
  Delabrouille, J.; Errard, J.; Hamilton, J.C.; Jones, M.; Krachmalnicoff, N.;
  Montier, L.; Natoli, P.; Piacentini, F.; Poletti, D.; Stompor, R.; Taylor,
  A.; Vielva, P.; Vittorio, N.
\newblock European Work on Future Ground-Based CMB Experiments.
\newblock {\em Bulletin of the AAS} {\bf 2019}, {\em 51}.
\newblock https://baas.aas.org/pub/2020n7i111.

\bibitem[Hazumi \em{et~al.}(2019)Hazumi et~al.]{LiteBIRD:2019}
Hazumi, M.; others.
\newblock LiteBIRD: A Satellite for the Studies of B-Mode Polarization and
  Inflation from Cosmic Background Radiation Detection.
\newblock {\em J Low Temp Phys} {\bf 2019}, {\em 194},~443–452.

\bibitem[Mukherjee \em{et~al.}(2020{\natexlab{a}})Mukherjee, Wandelt, and
  Silk]{Mukherjee:2019wfw}
Mukherjee, S.; Wandelt, B.D.; Silk, J.
\newblock {Multimessenger tests of gravity with weakly lensed gravitational
  waves}.
\newblock {\em Phys. Rev. D} {\bf 2020}, {\em 101},~103509,
  \href{http://xxx.lanl.gov/abs/1908.08950}{{\normalfont
  [arXiv:astro-ph.CO/1908.08950]}}.
\newblock
  doi:{\changeurlcolor{black}\href{https://doi.org/10.1103/PhysRevD.101.103509}{\detokenize{10.1103/PhysRevD.101.103509}}}.

\bibitem[Mukherjee \em{et~al.}(2020{\natexlab{b}})Mukherjee, Wandelt, and
  Silk]{Mukherjee:2019wcg}
Mukherjee, S.; Wandelt, B.D.; Silk, J.
\newblock {Probing the theory of gravity with gravitational lensing of
  gravitational waves and galaxy surveys}.
\newblock {\em Mon. Not. Roy. Astron. Soc.} {\bf 2020}, {\em 494},~1956--1970,
  \href{http://xxx.lanl.gov/abs/1908.08951}{{\normalfont
  [arXiv:astro-ph.CO/1908.08951]}}.
\newblock
  doi:{\changeurlcolor{black}\href{https://doi.org/10.1093/mnras/staa827}{\detokenize{10.1093/mnras/staa827}}}.

\bibitem[Chruslinska \em{et~al.}(2018)Chruslinska, Belczynski, Klencki, and
  Benacquista]{Chruslinska:2017odi}
Chruslinska, M.; Belczynski, K.; Klencki, J.; Benacquista, M.
\newblock {Double neutron stars: merger rates revisited}.
\newblock {\em Mon. Not. Roy. Astron. Soc.} {\bf 2018}, {\em 474},~2937--2958,
  \href{http://xxx.lanl.gov/abs/1708.07885}{{\normalfont
  [arXiv:astro-ph.HE/1708.07885]}}.
\newblock
  doi:{\changeurlcolor{black}\href{https://doi.org/10.1093/mnras/stx2923}{\detokenize{10.1093/mnras/stx2923}}}.

\bibitem[Chruslinska \em{et~al.}(2019)Chruslinska, Nelemans, and
  Belczynski]{Chruslinska:2018hrb}
Chruslinska, M.; Nelemans, G.; Belczynski, K.
\newblock {The influence of the distribution of cosmic star formation at
  different metallicities on the properties of merging double compact objects}.
\newblock {\em Mon. Not. Roy. Astron. Soc.} {\bf 2019}, {\em 482},~5012--5017,
  \href{http://xxx.lanl.gov/abs/1811.03565}{{\normalfont
  [arXiv:astro-ph.HE/1811.03565]}}.
\newblock
  doi:{\changeurlcolor{black}\href{https://doi.org/10.1093/mnras/sty3087}{\detokenize{10.1093/mnras/sty3087}}}.

\bibitem[Boco \em{et~al.}(2019)Boco, Lapi, Goswami, Perrotta, Baccigalupi, and
  Danese]{Boco:2019teq}
Boco, L.; Lapi, A.; Goswami, S.; Perrotta, F.; Baccigalupi, C.; Danese, L.
\newblock {Merging Rates of Compact Binaries in Galaxies: Perspectives for
  Gravitational Wave Detections}.
\newblock {\em The Astrophyscal Journal} {\bf 2019},
  \href{http://xxx.lanl.gov/abs/1907.06841}{{\normalfont
  [arXiv:astro-ph.GA/1907.06841]}}.
\newblock
  doi:{\changeurlcolor{black}\href{https://doi.org/10.3847/1538-4357/ab328e}{\detokenize{10.3847/1538-4357/ab328e}}}.

\bibitem[Boco \em{et~al.}(2021)Boco, Lapi, Chruslinska, Donevski, Sicilia, and
  Danese]{Boco:2020pgp}
Boco, L.; Lapi, A.; Chruslinska, M.; Donevski, D.; Sicilia, A.; Danese, L.
\newblock {Evolution of Galaxy Star Formation and Metallicity: Impact on Double
  Compact Objects Mergers}.
\newblock {\em Astrophys. J.} {\bf 2021}, {\em 907},~110,
  \href{http://xxx.lanl.gov/abs/2012.02800}{{\normalfont
  [arXiv:astro-ph.GA/2012.02800]}}.
\newblock
  doi:{\changeurlcolor{black}\href{https://doi.org/10.3847/1538-4357/abd3a0}{\detokenize{10.3847/1538-4357/abd3a0}}}.

\bibitem[Sathyaprakash \em{et~al.}(2012)Sathyaprakash
  et~al.]{Sathyaprakash:2012jk}
Sathyaprakash, B.; others.
\newblock {Scientific Objectives of Einstein Telescope}.
\newblock {\em Class. Quant. Grav.} {\bf 2012}, {\em 29},~124013,
  \href{http://xxx.lanl.gov/abs/1206.0331}{{\normalfont
  [arXiv:gr-qc/1206.0331]}}.
\newblock [Erratum: Class.Quant.Grav. 30, 079501 (2013)],
  doi:{\changeurlcolor{black}\href{https://doi.org/10.1088/0264-9381/29/12/124013}{\detokenize{10.1088/0264-9381/29/12/124013}}}.

\bibitem[Reitze \em{et~al.}(2019)Reitze et~al.]{Reitze:2019iox}
Reitze, D.; others.
\newblock {Cosmic Explorer: The U.S. Contribution to Gravitational-Wave
  Astronomy beyond LIGO}.
\newblock {\em Bull. Am. Astron. Soc.} {\bf 2019}, {\em 51},~035,
  \href{http://xxx.lanl.gov/abs/1907.04833}{{\normalfont
  [arXiv:astro-ph.IM/1907.04833]}}.

\bibitem[Namikawa \em{et~al.}(2022)Namikawa et~al.]{Namikawa:2021gyh}
Namikawa, T.; others.
\newblock {Simons Observatory: Constraining inflationary gravitational waves
  with multitracer B-mode delensing}.
\newblock {\em Phys. Rev. D} {\bf 2022}, {\em 105},~023511,
  \href{http://xxx.lanl.gov/abs/2110.09730}{{\normalfont
  [arXiv:astro-ph.CO/2110.09730]}}.
\newblock
  doi:{\changeurlcolor{black}\href{https://doi.org/10.1103/PhysRevD.105.023511}{\detokenize{10.1103/PhysRevD.105.023511}}}.

\bibitem[Aghanim \em{et~al.}(2020)Aghanim et~al.]{Aghanim:2018eyx}
Aghanim, N.; others.
\newblock {Planck 2018 results. VI. Cosmological parameters}.
\newblock {\em Astron. Astrophys.} {\bf 2020}, {\em 641},~A6,
  \href{http://xxx.lanl.gov/abs/1807.06209}{{\normalfont
  [arXiv:astro-ph.CO/1807.06209]}}.
\newblock
  doi:{\changeurlcolor{black}\href{https://doi.org/10.1051/0004-6361/201833910}{\detokenize{10.1051/0004-6361/201833910}}}.

\bibitem[Bianchini \em{et~al.}(2015)Bianchini, Bielewicz, Lapi, Gonzalez-Nuevo,
  Baccigalupi, Zotti, Danese, Bourne, Cooray, Dunne, Dye, Eales, Ivison,
  Maddox, Negrello, Scott, Smith, and Valiante]{Bianchini2015}
Bianchini, F.; Bielewicz, P.; Lapi, A.; Gonzalez-Nuevo, J.; Baccigalupi, C.;
  Zotti, G.D.; Danese, L.; Bourne, N.; Cooray, A.; Dunne, L.; Dye, S.; Eales,
  S.; Ivison, R.; Maddox, S.; Negrello, M.; Scott, D.; Smith, M.W.; Valiante,
  E.
\newblock Cross-correlation between the CMB lensing potential measured by
  Planck and high-z submillimeter galaxies detected by the Herschel-atlas
  survey.
\newblock {\em Astrophysical Journal} {\bf 2015}, {\em 802}.
\newblock
  doi:{\changeurlcolor{black}\href{https://doi.org/10.1088/0004-637X/802/1/64}{\detokenize{10.1088/0004-637X/802/1/64}}}.

\bibitem[Bianchini \em{et~al.}(2016)Bianchini, Lapi, Calabrese, Bielewicz,
  Gonzalez-Nuevo, Baccigalupi, Danese, de~Zotti, Bourne, Cooray, Dunne, Eales,
  and Valiante]{Bianchini2016}
Bianchini, F.; Lapi, A.; Calabrese, M.; Bielewicz, P.; Gonzalez-Nuevo, J.;
  Baccigalupi, C.; Danese, L.; de~Zotti, G.; Bourne, N.; Cooray, A.; Dunne, L.;
  Eales, S.; Valiante, E.
\newblock Toward a tomographic analysis of the cross-correlation between Planck
  CMB lensing and H-ATLAS galaxies.
\newblock {\em The Astrophysical Journal} {\bf 2016}, {\em 825},~24.
\newblock
  doi:{\changeurlcolor{black}\href{https://doi.org/10.3847/0004-637x/825/1/24}{\detokenize{10.3847/0004-637x/825/1/24}}}.

\bibitem[Abbott \em{et~al.}(2021{\natexlab{a}})Abbott
  et~al.]{LIGOScientific:2020kqk}
Abbott, R.; others.
\newblock {Population Properties of Compact Objects from the Second LIGO-Virgo
  Gravitational-Wave Transient Catalog}.
\newblock {\em Astrophys. J. Lett.} {\bf 2021}, {\em 913},~L7,
  \href{http://xxx.lanl.gov/abs/2010.14533}{{\normalfont
  [arXiv:astro-ph.HE/2010.14533]}}.
\newblock
  doi:{\changeurlcolor{black}\href{https://doi.org/10.3847/2041-8213/abe949}{\detokenize{10.3847/2041-8213/abe949}}}.

\bibitem[Abbott \em{et~al.}(2021{\natexlab{b}})Abbott
  et~al.]{LIGOScientific:2021qlt}
Abbott, R.; others.
\newblock {Observation of gravitational waves from two neutron star-black hole
  coalescences}.
\newblock {\em Astrophys. J. Lett.} {\bf 2021}, {\em 915},~L5,
  \href{http://xxx.lanl.gov/abs/2106.15163}{{\normalfont
  [arXiv:astro-ph.HE/2106.15163]}}.
\newblock
  doi:{\changeurlcolor{black}\href{https://doi.org/10.3847/2041-8213/ac082e}{\detokenize{10.3847/2041-8213/ac082e}}}.

\bibitem[Limber(1954)]{Limber:1954zz}
Limber, D.N.
\newblock {The Analysis of Counts of the Extragalactic Nebulae in Terms of a
  Fluctuating Density Field. II}.
\newblock {\em Astrophys. J.} {\bf 1954}, {\em 119},~655.
\newblock
  doi:{\changeurlcolor{black}\href{https://doi.org/10.1086/145870}{\detokenize{10.1086/145870}}}.

\bibitem[Lesgourgues(2011)]{lesgourgues2011cosmic}
Lesgourgues, J.
\newblock The Cosmic Linear Anisotropy Solving System (CLASS) I: Overview,
  2011,  \href{http://xxx.lanl.gov/abs/1104.2932}{{\normalfont
  [arXiv:astro-ph.IM/1104.2932]}}.

\bibitem[Blas \em{et~al.}(2011)Blas, Lesgourgues, and Tram]{Blas_2011}
Blas, D.; Lesgourgues, J.; Tram, T.
\newblock The Cosmic Linear Anisotropy Solving System (CLASS). Part II:
  Approximation schemes.
\newblock {\em Journal of Cosmology and Astroparticle Physics} {\bf 2011}, {\em
  2011},~034–034.
\newblock
  doi:{\changeurlcolor{black}\href{https://doi.org/10.1088/1475-7516/2011/07/034}{\detokenize{10.1088/1475-7516/2011/07/034}}}.

\bibitem[Smith \em{et~al.}(2003)Smith, Peacock, Jenkins, White, Frenk, Pearce,
  Thomas, Efstathiou, and Couchmann]{Smith:2002dz}
Smith, R.E.; Peacock, J.A.; Jenkins, A.; White, S.D.M.; Frenk, C.S.; Pearce,
  F.R.; Thomas, P.A.; Efstathiou, G.; Couchmann, H.M.P.
\newblock {Stable clustering, the halo model and nonlinear cosmological power
  spectra}.
\newblock {\em Mon. Not. Roy. Astron. Soc.} {\bf 2003}, {\em 341},~1311,
  \href{http://xxx.lanl.gov/abs/astro-ph/0207664}{{\normalfont
  [astro-ph/0207664]}}.
\newblock
  doi:{\changeurlcolor{black}\href{https://doi.org/10.1046/j.1365-8711.2003.06503.x}{\detokenize{10.1046/j.1365-8711.2003.06503.x}}}.

\bibitem[Ajith \em{et~al.}(2008)Ajith et~al.]{Ajith:2007kx}
Ajith, P.; others.
\newblock {A Template bank for gravitational waveforms from coalescing binary
  black holes. I. Non-spinning binaries}.
\newblock {\em Phys. Rev. D} {\bf 2008}, {\em 77},~104017,
  \href{http://xxx.lanl.gov/abs/0710.2335}{{\normalfont
  [arXiv:gr-qc/0710.2335]}}.
\newblock [Erratum: Phys.Rev.D 79, 129901 (2009)],
  doi:{\changeurlcolor{black}\href{https://doi.org/10.1103/PhysRevD.77.104017}{\detokenize{10.1103/PhysRevD.77.104017}}}.

\bibitem[Finn(1996)]{Finn:1995ah}
Finn, L.S.
\newblock {Binary inspiral, gravitational radiation, and cosmology}.
\newblock {\em Phys. Rev. D} {\bf 1996}, {\em 53},~2878--2894,
  \href{http://xxx.lanl.gov/abs/gr-qc/9601048}{{\normalfont [gr-qc/9601048]}}.
\newblock
  doi:{\changeurlcolor{black}\href{https://doi.org/10.1103/PhysRevD.53.2878}{\detokenize{10.1103/PhysRevD.53.2878}}}.

\bibitem[Taylor and Gair(2012)]{Taylor:2012db}
Taylor, S.R.; Gair, J.R.
\newblock {Cosmology with the lights off: standard sirens in the Einstein
  Telescope era}.
\newblock {\em Phys. Rev. D} {\bf 2012}, {\em 86},~023502,
  \href{http://xxx.lanl.gov/abs/1204.6739}{{\normalfont
  [arXiv:astro-ph.CO/1204.6739]}}.
\newblock
  doi:{\changeurlcolor{black}\href{https://doi.org/10.1103/PhysRevD.86.023502}{\detokenize{10.1103/PhysRevD.86.023502}}}.

\bibitem[Thrane and Romano(2013)]{Thrane:2013oya}
Thrane, E.; Romano, J.D.
\newblock {Sensitivity curves for searches for gravitational-wave backgrounds}.
\newblock {\em Phys. Rev. D} {\bf 2013}, {\em 88},~124032,
  \href{http://xxx.lanl.gov/abs/1310.5300}{{\normalfont
  [arXiv:astro-ph.IM/1310.5300]}}.
\newblock
  doi:{\changeurlcolor{black}\href{https://doi.org/10.1103/PhysRevD.88.124032}{\detokenize{10.1103/PhysRevD.88.124032}}}.

\bibitem[Aversa \em{et~al.}(2015)Aversa, Lapi, de~Zotti, Shankar, and
  Danese]{Aversa:2015bya}
Aversa, R.; Lapi, A.; de~Zotti, G.; Shankar, F.; Danese, L.
\newblock {Black Hole and Galaxy Coevolution from Continuity Equation and
  Abundance Matching}.
\newblock {\em Astrophys. J.} {\bf 2015}, {\em 810},~74,
  \href{http://xxx.lanl.gov/abs/1507.07318}{{\normalfont
  [arXiv:astro-ph.GA/1507.07318]}}.
\newblock
  doi:{\changeurlcolor{black}\href{https://doi.org/10.1088/0004-637X/810/1/74}{\detokenize{10.1088/0004-637X/810/1/74}}}.

\bibitem[Sheth \em{et~al.}(2001)Sheth, Mo, and Tormen]{Sheth:1999su}
Sheth, R.K.; Mo, H.; Tormen, G.
\newblock {Ellipsoidal collapse and an improved model for the number and
  spatial distribution of dark matter haloes}.
\newblock {\em Mon. Not. Roy. Astron. Soc.} {\bf 2001}, {\em 323},~1,
  \href{http://xxx.lanl.gov/abs/astro-ph/9907024}{{\normalfont
  [astro-ph/9907024]}}.
\newblock
  doi:{\changeurlcolor{black}\href{https://doi.org/10.1046/j.1365-8711.2001.04006.x}{\detokenize{10.1046/j.1365-8711.2001.04006.x}}}.

\bibitem[Lapi and Danese(2014)]{Lapi:2014ija}
Lapi, A.; Danese, L.
\newblock {Statistics of Dark Matter Halos in the Excursion Set Peak
  Framework}.
\newblock {\em JCAP} {\bf 2014}, {\em 07},~044,
  \href{http://xxx.lanl.gov/abs/1407.1137}{{\normalfont
  [arXiv:astro-ph.CO/1407.1137]}}.
\newblock
  doi:{\changeurlcolor{black}\href{https://doi.org/10.1088/1475-7516/2014/07/044}{\detokenize{10.1088/1475-7516/2014/07/044}}}.

\bibitem[Hui \em{et~al.}(2007)Hui, Gaztañaga, and LoVerde]{Hui_2007}
Hui, L.; Gaztañaga, E.; LoVerde, M.
\newblock Anisotropic magnification distortion of the 3D galaxy correlation. I.
  Real space.
\newblock {\em Physical Review D} {\bf 2007}, {\em 76}.
\newblock
  doi:{\changeurlcolor{black}\href{https://doi.org/10.1103/physrevd.76.103502}{\detokenize{10.1103/physrevd.76.103502}}}.

\bibitem[Scelfo \em{et~al.}(2018)Scelfo, Bellomo, Raccanelli, Matarrese, and
  Verde]{Scelfo:2018sny}
Scelfo, G.; Bellomo, N.; Raccanelli, A.; Matarrese, S.; Verde, L.
\newblock {GW$\times$LSS: chasing the progenitors of merging binary black
  holes}.
\newblock {\em JCAP} {\bf 2018}, {\em 09},~039,
  \href{http://xxx.lanl.gov/abs/1809.03528}{{\normalfont
  [arXiv:astro-ph.CO/1809.03528]}}.
\newblock
  doi:{\changeurlcolor{black}\href{https://doi.org/10.1088/1475-7516/2018/09/039}{\detokenize{10.1088/1475-7516/2018/09/039}}}.

\bibitem[Scelfo \em{et~al.}(2020)Scelfo, Boco, Lapi, and Viel]{Scelfo:2020jyw}
Scelfo, G.; Boco, L.; Lapi, A.; Viel, M.
\newblock {Exploring galaxies-gravitational waves cross-correlations as an
  astrophysical probe}.
\newblock {\em JCAP} {\bf 2020},
  \href{http://xxx.lanl.gov/abs/2007.08534}{{\normalfont
  [arXiv:astro-ph.CO/2007.08534]}}.

\bibitem[Zonca \em{et~al.}(2019)Zonca, Singer, Lenz, Reinecke, Rosset, Hivon,
  and Gorski]{Zonca2019}
Zonca, A.; Singer, L.; Lenz, D.; Reinecke, M.; Rosset, C.; Hivon, E.; Gorski,
  K.
\newblock healpy: equal area pixelization and spherical harmonics transforms
  for data on the sphere in Python.
\newblock {\em Journal of Open Source Software} {\bf 2019}, {\em 4},~1298.
\newblock
  doi:{\changeurlcolor{black}\href{https://doi.org/10.21105/joss.01298}{\detokenize{10.21105/joss.01298}}}.

\bibitem[{G{\'o}rski} \em{et~al.}(2005){G{\'o}rski}, {Hivon}, {Banday},
  {Wandelt}, {Hansen}, {Reinecke}, and {Bartelmann}]{2005ApJ...622..759G}
{G{\'o}rski}, K.M.; {Hivon}, E.; {Banday}, A.J.; {Wandelt}, B.D.; {Hansen},
  F.K.; {Reinecke}, M.; {Bartelmann}, M.
\newblock {HEALPix: A Framework for High-Resolution Discretization and Fast
  Analysis of Data Distributed on the Sphere}.
\newblock {\em ApJ} {\bf 2005}, {\em 622},~759--771,
  \href{http://xxx.lanl.gov/abs/arXiv:astro-ph/0409513}{{\normalfont
  [arXiv:astro-ph/0409513]}}.
\newblock
  doi:{\changeurlcolor{black}\href{https://doi.org/10.1086/427976}{\detokenize{10.1086/427976}}}.

\end{thebibliography}
\end{document}